# Persistent double layer formation in kesterite solar cells: a critical review

**Filipe Martinho[1], Simon Lopez-Marino[2], Moises Espíndola-Rodríguez[1,4], Alireza Hajijafarassar[2], Fredrik Stulen[3], Sigbjørn Grini[3], Max Döbeli[5], Mungunshagai Gansukh[1], Sara Engberg[1], Eugen Stamate[2], Lasse Vines[3], Jørgen Schou[1], Ole Hansen[2], Stela Canulescu[1]**

[1]Department of Photonics Engineering, Technical University of Denmark, DK-4000 Roskilde, Denmark.

[2]DTU Nanolab, Technical University of Denmark, DK-2800 Kgs. Lyngby, Denmark

[3]Department of Physics, University of Oslo, 0371 Oslo, Norway

[4]DTU Energy, Technical University of Denmark, DK-4000 Roskilde, Denmark.

[5]Ion Beam Physics, ETH Zurich, CH-8093 Zurich, Switzerland

Corresponding author: Filipe Martinho (filim@fotonik.dtu.dk), Stela Canulescu (stec@fotonik.dtu.dk)

## Abstract

In kesterite $Cu_2ZnSn(S,Se)_4$ (CZTSSe) solar cell research, an asymmetric crystallization profile is often obtained after annealing, resulting in a bilayered – or double-layered – CZTSSe absorber. So far, only segregated pieces of research exist to characterize the appearance of this double layer, its formation dynamics and its effect on the performance of devices. In this work, we review the existing research on double-layered kesterites and evaluate the different mechanisms proposed. Using a cosputtering-based approach, we show that the two layers can differ significantly in morphology, composition and optoelectronic properties, and complement the results with a large statistical dataset of over 850 individual CZTS solar cells. By reducing the absorber thickness from above 1000 nm to 300 nm, we show that the double layer segregation is alleviated. In turn, we see a progressive improvement in the device performance for lower thickness, which alone would be inconsistent with the well-known case of ultrathin CIGS solar cells. We therefore attribute the improvements to the reduced double-layer occurrence, and find that the double layer limits the efficiency of our devices to below 7%. By comparing the results with CZTS grown on monocrystalline Si substrates, without a native Na supply, we show that the alkali metal supply does not determine the double layer formation, but merely reduces the threshold for its occurrence. Instead, we propose that the main formation mechanism is the early migration of Cu to the surface during annealing and formation of $Cu_{2-x}S$ phases, in a self-regulating process akin to the Kirkendall effect. Finally, we comment on the generality of the mechanism proposed, by comparing our results to other synthesis routes, including our own in-house results from solution processing and pulsed laser deposition of sulfide- and oxide-based targets. We find that although the double layer occurrence largely depends on the kesterite synthesis route, the common factors determining the double layer occurrence appear to be the presence of metallic Cu and/or a chalcogen deficiency in the precursor matrix. We suggest that understanding the limitations imposed by the double layer dynamics could prove useful to pave the way for breaking the 13% efficiency barrier for this technology.

## 1 Introduction

Thin-film solar cells based on the kesterite $Cu_2ZnSn(S,Se)_4$ (CZTSSe) have conceptually been one of the most interesting areas of research in the field of photovoltaics over the last decades. To this day, it remains the only solar cell technology, besides crystalline Si, that is comprised of low-cost, earth-abundant and environmentally friendly elements, and that has surpassed 10% efficiency with long-term stability, without any encapsulation.[1,2] Moreover, just in the last 3 years alone there have been significant improvements in device optimization and fundamental understanding. At the synthesis level, new methods have been successfully implemented for high-efficiency kesterite synthesis, such as Molecular Beam Epitaxy (MBE)[3], and several optimizations have been achieved across traditionally reported methods, in particular sputtering,[4–6] including a recent record-tying certified 12.6%-efficient device, and an uncertified device with 13.0% efficiency[7]. A number of examples of strategies for back contact engineering and back surface passivation have been reported.[8–10] At the bulk level in kesterites, new insights have appeared on the coupling between O and Na,[11] on deep donor-acceptor defect centers[12] and the origin of loss mechanisms, namely the open-circuit voltage deficit ($V_{oc}$ deficit), now widely regarded as the main bottleneck in kesterite solar cells.[13–16] Many efforts have been put recently on cationic substitution or alloying (loosely referred to as doping in this field), in particular with Ag, Ge, and alkali metals,[17–22] but also other elements such as Ba or Sr,[23,24] further revealing insights on bulk properties and performance limitations. At the heterojunction interface level, new passivation techniques have been discovered, further reducing the $V_{oc}$ deficit component associated with interface recombination.[25–27] Also in this regard, perhaps the biggest recent breakthrough in the kesterite field has been the replacement of CdS as n-type buffer layer in kesterite solar cells, and the demonstration of Cd-free devices with efficiency above or at parity with CdS, in particular for the pure sulfide (CZTS) kesterite.[28–30] The importance of this result is that it further highlights that the current bottleneck seems to be not at the interface level, but due to shortcomings in the bulk properties of kesterites, namely band tailing, recombination through deep defects and very low bulk carrier lifetimes and diffusion lengths overall.[31] Despite the recent notable achievements mentioned above, the certified record efficiency of all kesterite solar cells has remained at 12.6% for nearly a decade now.[7,32] Among the major macrophysical experimental constraints identified by researchers, inherent difficulties in controlling the phase purity, composition, growth conditions and reproducibility are often appointed as the dominant limitations for experimentalists in the field.[13,33–35] Examples which result from these non-ideal conditions, quite often seen in literature, are the reports of kesterite decomposition, blistering occurrence and back contact degradation issues, all with possible corresponding adhesion issues.[27,36–40] Another example is an asymmetric crystallization profile during the sulfurization step, which manifests itself through the appearance of a bilayer, or double layer, in the kesterite absorber. Throughout our work on the sulfide kesterite (CZTS) from cosputtered Cu, SnS and ZnS precursors, we have found the occurrence of double-layered absorbers to be extremely difficult to circumvent, and persistently occurring under almost every relevant processing parameter. Surprisingly, unlike the other examples mentioned above, we have found specific studies on this double layer phenomenon to be lacking, despite a large number of occurrences throughout the literature, as will be discussed below. Therefore, in this contribution we attempt to address this problem in detail. We discuss the double layer formation mechanism, the influence of the different processing parameters, and present depth-resolved characterization of double-layered CZTS absorbers, both on Mo-coated soda lime glass (SLG) and on Si wafer-based substrates. We complement the results with large statistical datasets from over 850 finished solar cells to confirm that the different processing parameters are relevant within the space of device optimization. Furthermore, we discuss the impact of the kesterite synthesis route on the double layer dynamics and present a direct comparison of this effect for absorbers grown in-house using three different synthesis methods: cosputtering, pulsed laser deposition (PLD) and solution processing. Essentially, the results show that for our cosputtering approach, the double layer occurrence places an unfavorable upper limit on the ideal absorber thickness of about 600 nm. While these thin absorbers achieve the highest-performing devices with up to 6.82% efficiency, we observe a clear degradation in device parameters for increasing absorber thicknesses up to 1 µm, in particular in the fill factor (FF), series resistance $R_S$, shunt resistance $R_{sh}$, diode ideality factor $A$ and reverse saturation current density $J_0$, which we attribute to the effects of the double-layer presence.

This work is organized as follows: in **Section 2**, we illustrate examples of double-layered kesterites and review the existing literature where a bilayer or double layer could be identified in the kesterite absorber. This section aims to provide the reader without any *a priori* knowledge an understanding on the current aspects being discussed in literature regarding double-layered kesterites. In the results section, **Section 4**, we present detailed experimental results on the double layer occurrence in our cosputtering system, with the following subsections:

4.1. Double layer formation mechanism
4.2. Characterization of double-layered absorbers from a two-stage annealing process
4.3. Influence of annealing parameters
4.4. Influence of absorber thickness
4.5. Effects of postdeposition heat treatments in double-layered absorbers
4.6. Influence of alkali metal availability (growth on Si substrates)
4.7. Comments on the generality of the double layer formation mechanism proposed in this work

## 2 Survey of occurrence of double-layered kesterites in literature

A double-layered kesterite absorber can be clearly identified in a cross-section scanning electron microscope (SEM) image of an absorber or a finished solar cell, as shown in **Figure 1**. These correspond to typical examples from vacuum processing methods (cosputtering and PLD) and solution processing in our group. In some cases, the two layers are clearly separated by a nearly flat interface, as shown in **Figure 1 (a)**. More often, the top CZTS layer exhibits large crystallites, while the bottom consists of a thin layer with small grains and possibly secondary phases, as in **Figure 1 (b)**. In solution processing, a double layer often appears using a precursor solution of metal salts, as shown in **Figure 1 (c)**. Also from the solution processing route, a more extreme case, shown in **Figure 1 (d),** can occur in solution-processed kesterites from the nanoparticle synthesis route where a large accumulation of carbon residues are known to occur at the back.[41] A fifth case, shown in **Figure 1 (e)**, can be found in alternative kesterite synthesis routes based on oxide precursors. The example of **Figure 1 (e)** corresponds to a CZTS device based on pulsed laser deposition from an oxide target, which we have discussed in previous work.[42]

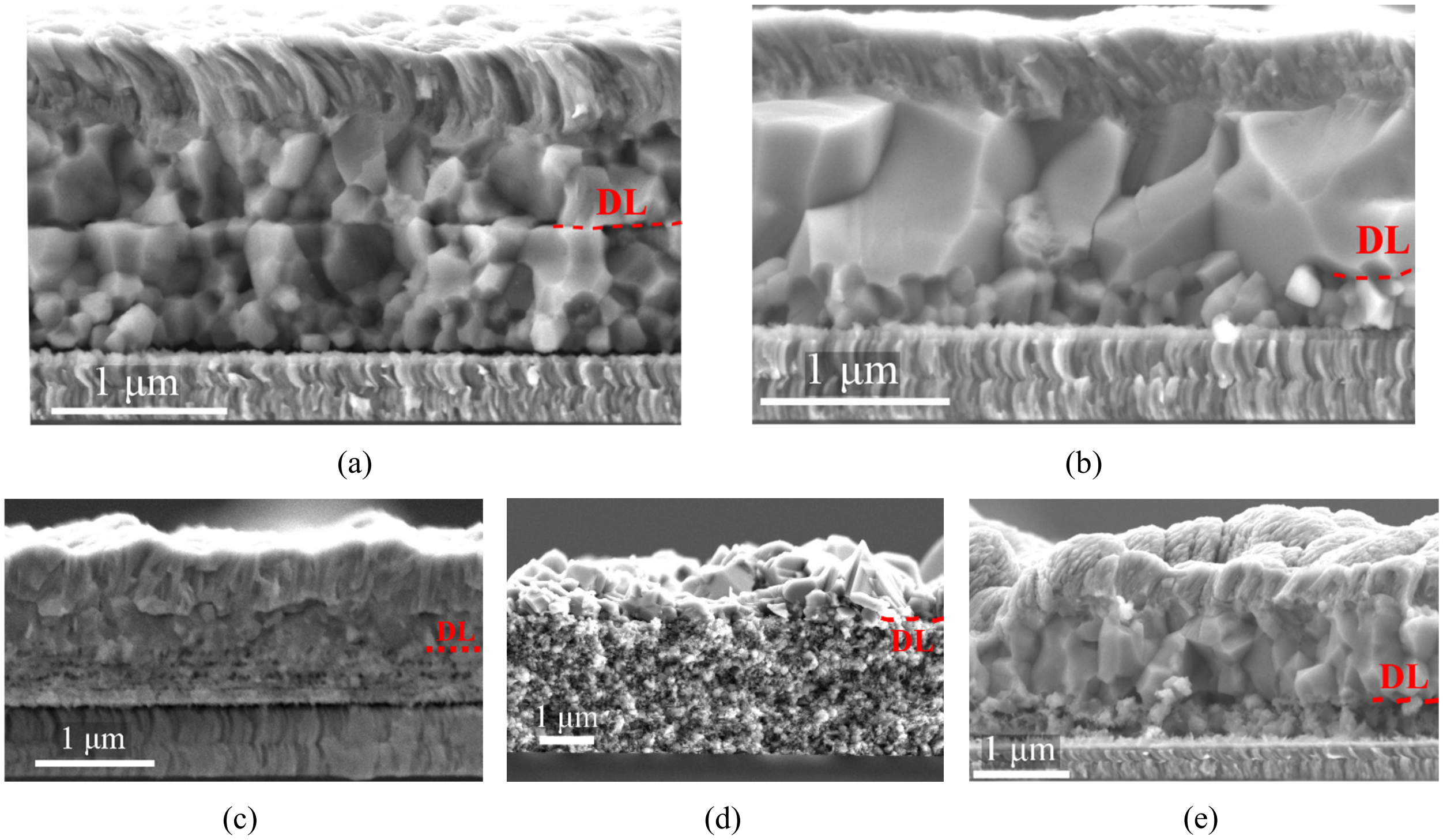

**Figure 1** – Typical SEM cross-section images of CZTS solar cells and absorbers exhibiting a double layer structure. (a) and (b) are full devices made from vacuum-processed precursors deposited using cosputtering; (c) is a full device from solution processing of metal salts; (d) is an absorber from solution processing of nanoparticles, (e) is a CZTS device from pulsed laser deposition of a single oxide target. In (a), (b), (c) and (e) the solar cell structure consists of Mo/CZTS/CdS/i-ZnO/Al:ZnO, from bottom to top. The red dashed line marks the double layer (DL) interface.

In this work, we will discuss mostly the cases of **Figure 1 (a)**, **(b)** and **(c)**. There is a number of instances of this kind of double layer appearing in literature, and important direct or indirect insights on its dynamics appear across multiple publications. A summary of the most critical aspects found is compiled in **Table 1**. Chalapathy et al. identified a double layer morphology in CZTSSe absorbers from stacked metallic precursors deposited by sputtering (Mo/Zn/Cu/Sn/Cu), and noted that an increased Cu content and modified stacking order improved the morphology (i.e. reduced the fine grain layer at the bottom).[43] Zhang et al., on two different occasions, observed a double layer in CZTS absorbers also from sputtered metallic precursors, with different stacking orders (Mo/Zn/Sn/Cu and Mo/Zn/Sn/Cu/Zn), and suggested an improvement in performance due to a Zn enrichment of the surface.[44,45] There, the double layer morphology was visible in both configurations, although the stacking order clearly influenced the final absorber morphology. Cao et al. reported a 9.6% efficient cell from a double-layered CZTSSe absorber from spin-coating of precursors consisting of nanoparticle-containing inks.[41] As mentioned above, in that work the double-layer morphology obtained was attributed to an accumulation of carbon on the backside, and the authors suggested that a better control of the double-layer formation could lead to further device improvements. A similar result was obtained by Zhou et al. using nanoparticles with Na inclusion.[46] Gang et al. studied the influence of a "soft-annealing" (a pre-annealing in an inert atmosphere, at a temperature lower than the actual sulfurization/selenization annealing) on CZTSSe solar cells, using a sputtered stack of metals (Mo/Cu/Sn/Zn). The pre-annealing appears to influence the final morphology of the kesterite absorbers, and a double-layer configuration is seen in some cases, but it was not discussed by the authors.[47] In a series of five publications, Kim and coworkers mentioned that there is currently not enough data to clarify the origin of the double layer formation in CZTSSe kesterites.[4,7,9,48,49] In these works, the authors addressed this issue with very detailed and comprehensive formation studies, revealing that the double layer appears from the early stages of the annealing process, but the exact mechanism seems to be dependent on the precursor configuration. Using a sputtered metal stack of Mo/Zn/Cu/Sn, Kim et al. achieved a 12.5% efficient CZTSSe device but found it to exhibit a clear double layer, with a uniform upper layer with large grains and a bottom layer consisting of a mix of voids and smaller CZTSSe grains and Zn-SSe, Cu-Sn-SSe and Cu-SSe secondary phases.[4] The authors suggest that removing the voids and secondary phases could improve the FF and short circuit current density ($J_{sc}$). Through a low temperature formation study, the authors show that the double layer arises during sulfo-selenization due to a complex combination of a formation of a ZnSSe layer by dezincification of the Cu-Zn precursor alloy, followed by a mass transfer of Cu and Sn. In a subsequent publication using the same Mo/Zn/Cu/Sn precursor configuration, the same group further shows that the secondary phase formation associated with the double layer is caused by a good wettability between liquid Sn and Mo, causing a persistent formation of Cu-SSe and Cu-Sn-SSe phases near the back contact.[9] By placing a 5 nm $Al_2O_3$ layer between CZTSSe and Mo, the Sn wetting was reduced and the double layer visibly improved. In another publication, by further fine-tuning of the experimental parameters, the same group achieved a certified 12.62% CZTSSe device.[7] In that work, Son et al. compared a pure selenization and a sulfo-selenization of the same metal precursor stack for different annealing temperatures and steps. As in the previous works, the double layer is not only visible at these intermediate stages, but also in the final high-efficiency devices. Interestingly, despite the undesirable double layer, Son et al. mention the possibility of an unintentional local contact structure, with the lower CZTSSe bridging to the back contact through the voids and a reduced interface recombination due to a passivating effect of the ZnSSe. When using instead a sputtered Mo/ZnS/SnS/Cu configuration, with a selenization[49] or a sulfo-selenization[48] annealing approach, the same group showed that a uniform single layer CZTSSe can be obtained, highlighting the importance of the precursor configuration. Nevertheless, the double-layer dynamics could still be seen at the intermediate annealing steps, in particular between 330 and 420 °C where a CuSe layer with large grains forms at the surface. Although it is not discussed directly in the aforementioned series, we note that it seems that in these works the chalcogen introduction approach during annealing – whether from S or Se powders/shots, from $H_2S$ or $H_2Se$ or from $SeS_2$ – could also lead to different double layer dynamics. Possible clues are, for instance, the different reactivity of these chalcogen sources and the thickness of the $Mo(S,Se)_2$, an indirect indicator of the penetration of the chalcogen during annealing. Moreover, direct comparisons between sulfurization and selenization have shown that sulfurized films (CZTS) tend to show a larger variance in domain size, notably with larger domains at the surface, whereas selenized films (CZTSe) exhibit more uniform domain sizes across the depth.[50] These could be indications that the selenide kesterite system may be inherently less prone to the occurrence of a double layer compared to the sulfide. Another work directly studying the double layer dynamics and its impact on device performance in kesterites was conducted by Zhang et al..[51] There, metal precursors were deposited by electrodeposition with a stack of Mo/Cu/Sn/Zn, and subsequently pre-heated at 250 °C in an inert atmosphere, pre-selenized from 250 to 350 °C and finally selenized at 550 °C. From that synthesis process resulted a clear double-layered CZTSe absorber, which the authors assigned to an insufficient reaction between ZnSe and the Cu-Sn-Se phases near the bottom Mo contact. Another important detail in that study was the absence of any $MoSe_2$ formation. The authors found the double layer distribution to hinder the solar cell performance by causing a back-contact barrier with high series resistance. By tuning the kesterite composition with an increased Sn amount in the precursor stack, Zhang et al. were able to completely eliminate the double layer distribution, and attributed the improvement to a facilitated diffusion of the liquid $Cu_2Se$ phase. In this work, we will show that this result is only a small aspect of a more general double layer dynamics, where there are other parameters at play. For instance, Haass et al. studied the alkali doping of CZTSSe from spin-coating of a precursor solution containing metal chloride salts in dimethyl sulfoxide (DMSO), using a $SiO_x$ alkali diffusion barrier between the SLG and the Mo.[17] In their work, the reference sample (with no alkali supply) shows a clear double layer distribution. The authors specifically analyze the evolution of the double layer dynamics for different Sn contents in their reference sample and find that the Sn has no effect on the double layer, in direct contradiction with the findings of the previous study mentioned above. In fact, Haass et al. proved that the different alkali metals themselves play a crucial role as fluxing (or surfactant) agents in the crystallization and final morphology of the kesterite absorber, resulting in a 12.3% efficient solar cell with a single layer of coarse grains. This solution processing approach is similar to the one we described in **Figure 1 (c)**. For this synthesis route, the solvent drying step and the presence of organic compounds such as Thiourea and DMSO are potentially additional aspects to consider regarding the double layer occurrence.

**Table 1** – Summary of the survey of double-layered kesterite occurrences in literature.

| Type of work | Precursor | Annealing | Double layer (DL) findings and comments | Ref. |
|---|---|---|---|---|
| **Standard CZTSSe** | Mo/Zn/Cu/Sn/Cu (sputtering) | Pre-annealing 300 °C 1 h in Ar + sulfo-selenized 545 °C 10 min 10 °C/min 400 Torr | Cu content and stacking order influence DL; Composition gradient. 8.03% CZTSSe | 43 |
| | Mo/Zn/Sn/Cu & Mo/Zn/Sn/Cu/Zn (sputtering) | Sulfurized 570 °C 30 min 20 °C/min 0.45 atm | Stacking order influences DL. 5.55% CZTS | 44,45 |
| | Binary and ternary sulfide nanoparticles (solution processing) | N.A. | C-residues accumulated at the back. 9.6% CZTSSe | 41,46 |
| | Mo/Cu/Sn/Zn (sputtering) | Pre-annealing in $N_2$ + sulfo-selenized 540 °C 10 min | No discussion in the text. Possible influence of pre-annealing. 9.24% CZTSSe | 47 |
| | Mo/ZnS/SnS/Cu (sputtering) | Selenized 250-590 °C 20 min, 1 atm | DL visible between 350-420 °C. Possible CuSe surface formation causing DL. 9.47% CZTSSe | 48,49 |
| | Mo/Zn/Cu/Sn (sputtering) | Sulfo-selenized in $H_2S$ + Se powder 480 °C 10 min, 700 torr | DL visible in intermediate steps and final devices. Good Sn wetting on Mo and Zn(S,Se) at the back cause DL, possibly beneficial. 12.62% CZTSSe | 4,7,9 |
| | Mo/Cu/Sn/Zn (electrodeposition) | Pre-annealing 250 °C 120 min in $N_2$ + pre-selenized 250 to 350 °C + selenized 550 °C 20 min 40 °C/min with Se (70 Pa) and $SnSe_x$ (1 Pa) | DL assigned to insufficient ZnSe and Cu-Sn-Se reaction. DL causes back contact barrier with high $R_s$. Increase in Sn promotes CuSe diffusion, solves DL. 7.2% CZTSSe | 51 |
| **Alkali doping & cation substitution** | Metal salts in DMSO (spin-coating) | 3-stage selenization (300, 500 and 550 °C) | Alkali metal supply improves DL. Sn content has no effect on DL. 12.3% CZTSSe | 17 |
| | Mo/Cu/Sn/Cu/Zn (sputtering) + Ge (evaporation), multiple studies | Sulfurized 200 °C 45 min + 550 °C 5 min 20 °C/min (1 bar) | Ge alloying has low impact on DL (no fluxing agent behavior). 5.6% Ge:CZTS | 22 |
| | | Selenized 400 °C 30 min + 550 °C 15 min 20 °C/min (1 bar) | Ge can aggravate or eliminate DL depending on location ($GeSe_2$ fluxing agent). DL connected to out-diffusion of Cu to surface and $Cu_xSe$ formation. DL shows composition difference. 11.8% Ge:CZTSe | 20,21,52 |
| | | Selenized 400 °C 30 min + 550 °C 15 min 20 °C/min (1 bar) | The two layers in the DL have different composition and different grain boundaries | 53 |
| **Alternative configurations & substrates** | Metal salts in DMSO (doctor-blade coating on TiN/Mo) | Selenized 550 °C 20 min | TiN and Mo influence DL. A triple-layered kesterite can appear. Asymmetric Na distribution. 7.1% CZTSSe (median) | 8 |
| | Mo/Ti/Cu-Sn-ZnS (cosputtering) | Sulfurized 550 °C 60 min | No discussion. Clear DL even in reference. | 54 |
| | Mo foil/Zn/Sn/Cu (sputtering) | Pre-annealing 300 °C 1 h in Ar + sulfo-selenized 480-540 °C 7.5 min 10 °C/min | Suggestion of alkali inclusion to improve DL. 8% CZTSSe on flexible substrate | 6 |
| | FTO/(Mo)/Cu/Sn/Cu/Zn (sputtering) + Ge (evaporation) | Selenized 400 °C 30 min + 525 or 550 °C 15 min 20 °C/min 1 bar | Mo and Mo:Na reduce DL formation. 7.7% bifacial CZTSSe | 55 |
| | 2 sputtering runs | 2 annealing runs | Intentional DL, composition gradient impacts devices. 8.8% CZTS | 56,57 |

Additionally, there have been double layer occurrences in studies on germanium doping (or alloying). Sanchez et al. studied the Ge doping of the sulfide kesterite,[22] and Giraldo et al. and Thersleff et al. studied the Ge doping of the selenide kesterite on multiple works,[20,21,52,53] where the kesterite base was obtained by sulfurization or selenization of a sputtered metallic precursor stack of Mo/Cu/Sn/Cu/Zn, respectively. In all cases, double-layered absorbers appear, even on reference non-doped samples, and important insights on the double layer dynamics are indirectly reported. Giraldo et al. report that an unfavorable "bilayer" distribution is often seen in standard CZTSSe absorbers, and suggest that Ge doping can play an important role in promoting a complete crystallization of the kesterite.[21] Remarkably, it was found in these studies that a Ge layer only a few nm thick could either significantly aggravate or completely eliminate the double layer distribution, depending on its position in the precursor stack.[52] The result was attributed to the fluxing agent effect of the $GeSe_2$ phase, which has a low melting point around 380 °C.[20] Sanchez et al. noted that for sulfide kesterites, no such Ge-S phase with low melting point exists, and therefore Ge doping has a lesser impact on the crystallization dynamics in CZTS.[22] Through a very detailed formation study at various temperatures and annealing stages, Giraldo et al. noted that for undoped CZTSe absorbers, the double layer distribution appeared to be connected to an out-diffusion of Cu to the surface.[21] There, the liquid $Cu_xSe$ phases would promote the formation of large grains near the top, and voids and small grains would occur near the bottom of the kesterite. Furthermore, the authors identified a large compositional difference between the two layers, which remained after the full processing. Notably, it was found that a small amount of Ge significantly reduced this out-diffusion of Cu and enhanced the crystallization and composition uniformity, which ultimately resulted in a 11.8%-efficient Ge-doped CZTSe cell.[21] Intriguingly, a detailed STEM-EELS analysis of a 10.1%-efficient Ge-doped CZTSe device by Thersleff et al. revealed that even in Ge-doped CZTSe some aspects of a double-layer distribution still remain. Namely, a different stoichiometry in the upper and lower part of the kesterite, and a remarkable difference in the corresponding grain-boundaries were detected, with different chemical composition and a presence of different elements and phases.[53]

Finally, multiple double layered kesterites have appeared in reports on alternative substrate configurations. Using a precursor solution of metal salts in DMSO, Schnabel and Ahlswede tested the use of a TiN barrier at the Mo/kesterite interface to inhibit the degradation of the back-contact interface.[8] Although a CZTSSe absorber with a double layer appeared in all samples, that work showed that the TiN changed the crystallization dynamics, in particular near the back contact. Interestingly, when the kesterite is in direct contact with Mo (Mo/CZTS reference and Mo/TiN/Mo/CZTS), a layer with larger grains appeared near the back, effectively creating a triple-layered kesterite absorber. In addition, a very thick (1 μm) $Mo(S,Se)_x$ was formed in the reference sample despite the double layer formation, indicating full penetration of the chalcogen during annealing. Other notable features were a relatively uniform composition with depth, but an asymmetric distribution of Na in the absorber layer with the addition of TiN compared to the reference. Guo et al. studied the effect of a thin Ti layer at the back interface in CZTS cells fabricated by cosputtering of Cu, Sn and ZnS.[54] Although a double layer appeared in all cases, including the reference, no particular aspects of the double layer formation were discussed by the authors. Jo et al. produced CZTSSe devices on Mo foil substrates from a sputtered stack of metals (Mo/Zn/Sn/Cu).[6] The precursors were then "soft"-annealed (pre-annealed) in an inert atmosphere at 300 °C for 1 h, and then sulfo-selenized between 480 and 540 °C. In the final device, a clear double layer morphology was visible, and although the authors identified the occurrence, there was only a suggestion that alkali doping could provide some future improvements. Espindola-Rodriguez et al. produced bifacial CZTSe and CZTSSe cells on FTO substrates, using a similar sputtered metallic stack of Cu/Sn/Cu/Zn as mentioned above.[55] They found that the reference FTO/kesterite case would exhibit a double-layer morphology, with large grains at the top and small grains at the bottom (see, in particular, the supporting information of that work).[55] Additionally, it was revealed that both the introduction of a thin Mo layer or a Na-doped Mo layer at the FTO/kesterite interface significantly improved the absorber morphology, reducing the double-layer distribution. These results further highlight the complicated influence of alkali metals and the Mo/kesterite interface on the double layer dynamics. Lastly, it is worth mentioning the works of Tajima et al.[56] and Yan et al.,[57] where the researchers made intentional use of a double layer to try to improve the properties of CZTS kesterites, by specifically tuning the Cu/Sn ratio in each layer. Differently from all the above-mentioned works, both Tajima et al. and Yan et al. achieved the double layer kesterite using two different sputtering and two different annealing runs per sample. In both cases, it was found that the solar cells from double-layered CZTS exhibited significantly different performance parameters compared to the (single layer) reference, hinting that the composition gradients existing in the double layer can affect the devices.

## 3 Experimental Section

The CZTS devices produced in this work were obtained from a cosputtering route. The precursor films were obtained from cosputtering of Cu, SnS and ZnS targets with a diameter of 2 in, by using a DC power supply for Cu and RF power supplies for SnS and ZnS. The targets were positioned at an angle of 63° between the target surface normal and the substrate plane. The target to substrate distance was kept at 7 cm, and the deposition pressure was $4\times10^{-3}$ mbar unless otherwise specified. No intentional substrate heating was used. The composition of the precursor films was varied by tuning the power density applied to the targets, in the ranges of 0.48-0.58 W/in$^2$ for Cu, 2.63-2.94 W/in$^2$ for SnS and 3.42-4.62 W/in$^2$ for ZnS. The composition of the films was estimated by Energy Dispersive X-ray Spectroscopy (EDS), except for films grown on monocrystalline Si substrates, where the composition of both the precursor and annealed absorber was measured by Rutherford Backscattering Spectrometry (RBS). The thickness of the precursor films was controlled by changing the deposition time, and calibrated using SEM cross-section images of the final absorber after annealing. The precursors were then annealed in a sulfur- and tin sulfide-containing atmosphere. In this work, we present results from two different annealing tools. The annealing parameters used in both tools are detailed in **Table 2**. The annealing furnace used for the two-stage (2S) process consists of a semi-open quartz tube inserted into a muffle oven with a very high thermal inertia, for which only very slow heating and cooling profiles are possible. The annealing furnace used for the one-stage (1S) process consists of an open quartz tube furnace with a 3-zone PID control from Hobersal. In this furnace, the annealed samples can be rapidly quenched by opening the furnace latches. The base pressure for both furnaces was in all cases lower than $1\times10^{-2}$ mbar. The initial pressure was set in both furnaces by supplying nitrogen gas using an injection butterfly valve. After annealing, the kesterite absorbers were chemically etched in a 10 wt% ammonium sulfide ($(NH_4)_2S$) solution for 5 min. Then, a 50 nm CdS buffer layer was deposited by chemical bath deposition using cadmium sulfate ($CdSO_4$) as cadmium source, thiourea ($CH_4N_2S$) as sulfur source, ammonium hydroxide ($NH_4OH$) as complexing agent and ammonium chloride ($NH_4Cl$) as buffer solution. The deposition was done at a constant temperature of 70 °C. Following that, 50 nm intrinsic ZnO (i-ZnO) and 350 nm Al-doped ZnO (AZO) layers were deposited by RF sputtering. During the AZO sputtering, the substrate temperature was kept at 150 °C. The finished solar cells were then mechanically scribed to define a nominal cell area of 0.09 $cm^2$ (3×3 $mm^2$). The exact area of the cells was measured using the software ImageJ 1.47t. Neither antireflection coating nor metallic grid were used in this work. For the samples with intentional Na doping, the doping of the samples was achieved by thermal evaporation of NaF.

The Secondary Ion Mass Spectrometry (SIMS) depth profiles were obtained from a Cameca IMS-7f microprobe. A 5 keV $Cs^+$ primary beam was employed and rastered over 150×150 $\mu m^2$, and the positive ions were collected from a circular area with a diameter of 33 μm. The RBS measurements were done using 2 MeV He ions and a silicon PIN diode detector under a 168° angle. The collected RBS data were analyzed and fitted using the software RUMP,[58] from which the depth-resolved composition of the precursor and annealed films on monocrystalline Si substrates was estimated. Details on the fabrication of the Si structures used in SIMS and RBS can be found in previous work.[59,60]

The Raman spectroscopy measurements were done on absorber layers or complete cells with an excitation wavelength of 785 and 532 nm, using a modified Renishaw Raman spectrometer equipped with a Si CCD detector, in confocal mode using a 50× objective lens. For the 532 nm laser, a power of 0.39 mW was used, with a laser spot size of 0.9 μm, whereas for the 785 nm laser a power of $5.4\times10^{-5}$ mW was used, with a spot size of 1.3 μm. In both cases, the exposure time used was 10 s, resulting in a total of 10 acquisitions averaged per spectrum.

The I–V characteristic curves of the solar cells were measured at near Standard Test Conditions (STC: 1000 $W/m^2$, AM1.5 and 25 °C). A Newport class ABA steady state solar simulator was used. The irradiance was measured with a 2×2 $cm^2$ Mono-Si reference cell from ReRa certified at STC by the Nijmegen PV measurement facility. The temperature was kept at 25 ± 3 °C, as measured by a temperature probe on the sample holder. The acquisition was done with 2 ms between points, using a 4 wire measurement probe, from reverse to forward voltage, acquiring a total of 150 points per IV curve. The external quantum efficiency (EQE) was measured using a QEXL setup (PV Measurements) equipped with a grating monochromator, adjustable bias voltage, and a bias spectrum.

The scanning electron microscopy (SEM) images shown in this work were acquired using a Zeiss Merlin field emission electron microscope under a 5 kV acceleration voltage.

The X-ray Diffraction (XRD) patterns were obtained for CZTS on SLG/Mo and CZTS on monocrystalline Si substrates using a grazing incidence mode with incidence angles from 1.5° to 6°. The step size was 0.02°, with an acquisition time of 5 s per step. The X-ray source was from Cu Kα radiation. A 0.2 mm slit was positioned in front of the source to get a narrow beam, and a parabolic mirror was used in the detector for grazing incidence measurements.

## 4 Results and Discussion

In this work, the CZTS absorber layers were produced in two different annealing tools, using a two-stage (2S) process and a one-stage (1S) process. The reference annealing conditions for each process are described in **Table 2**. Several variations of these parameters are studied in this work, and will be described case-by-case.

**Table 2** – Reference annealing conditions used for device fabrication in this work. For the dwell conditions, the pressure value was either forced (F) using a valve, or a result of gas expansion (G). For the 2S process, the 1.5 mbar pressure was dynamically kept until the end of the first stage.

| Annealing type | Initial conditions | Heating rate 1 (°C/min) | Dwell conditions stage 1 | Heating rate 2 (°C/min) | Dwell conditions stage 2 | Cooling |
|---|---|---|---|---|---|---|
| Two-stage (2S) | 1.5 mbar, 100 mg S, 10 mg Sn | 10 | 200 °C, 30 min, then 660 mbar (F) | 10 | 575 °C, 30 min, 1000 mbar (G) | Very slow (~ 8 hours) |
| One-stage (1S) | 175 mbar, 50 mg S, 5 mg Sn | 20 | 575 °C, 45 min, 300 mbar (G) | - | - | Quench at 300 °C |

### *4.1 Double layer formation mechanism*

We assess the double layer formation mechanism by analyzing samples produced at a low temperature annealing, in the range 200-400 °C, which would correspond to the first stage in a two-stage (2S) annealing. In **Figure 2**, the resulting morphologies are compared to the initial as-sputtered precursor (**Figure 2 (a)** and **(b)**). The precursor exhibits a uniform morphology consisting of a nearly amorphous matrix of Cu, SnS and ZnS. The precursor composition was estimated using EDS as Cu/Sn = 1.86, Zn/Sn = 1.10 and S/metals = 0.66. After the first annealing stage at temperatures as low as 200 °C, double-layer features become evident, as shown in **Figure 2 (c)-(g)**, with the appearance of a large-grained top layer.

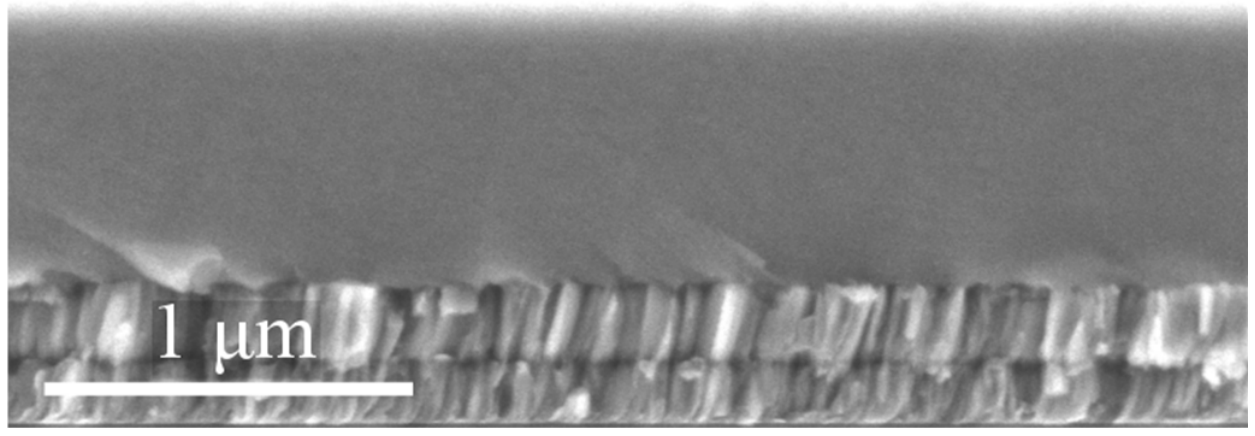

(a) Precursor, cross-section

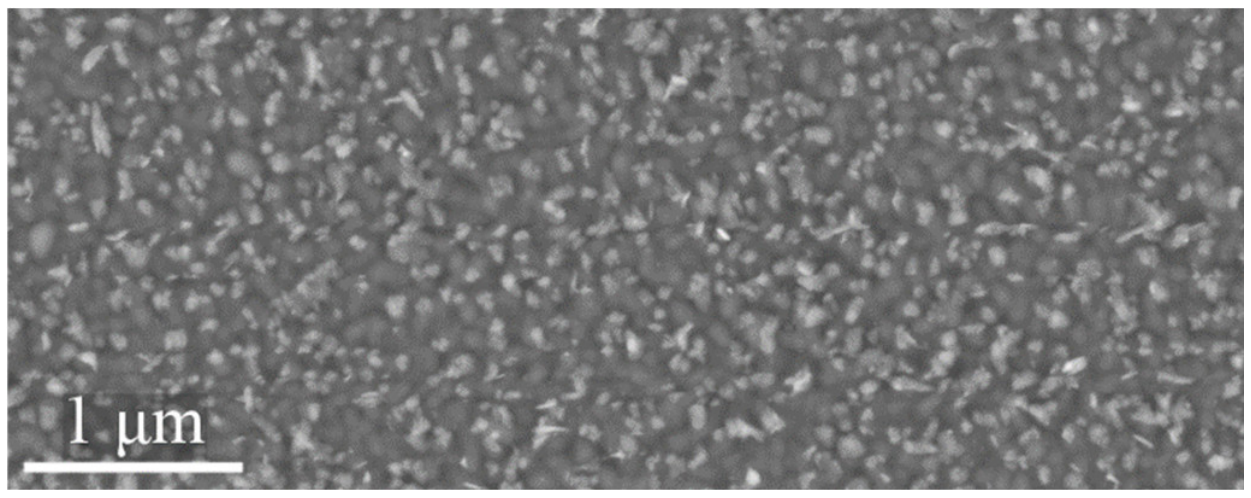

(b) Precursor, top view

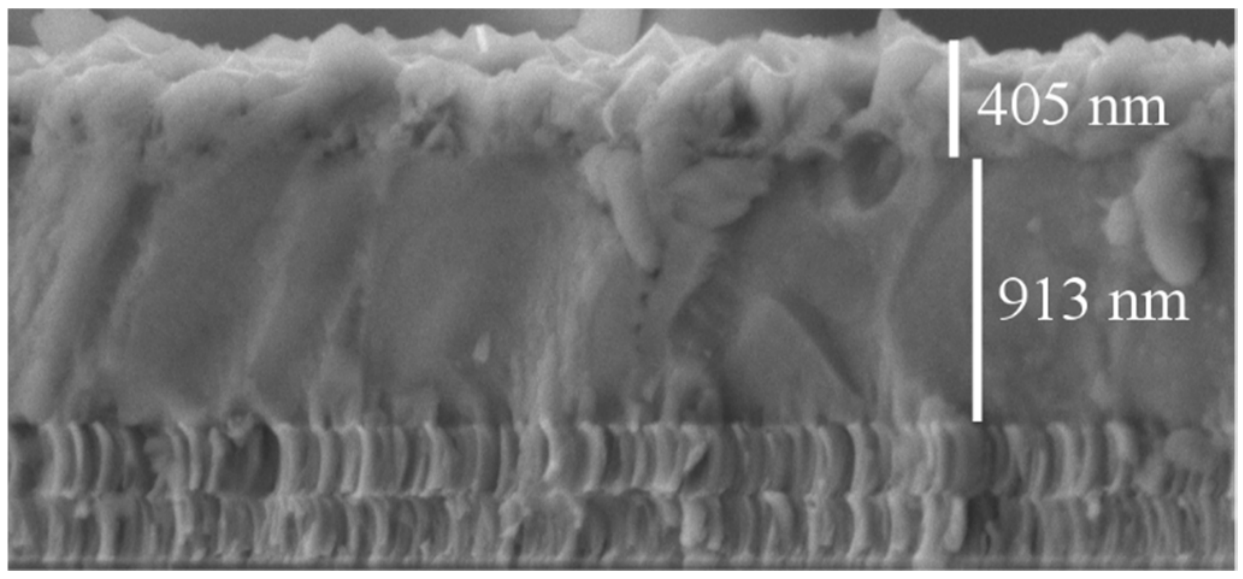

(c) 2S-Reference: 200 °C 30 min 1.5 mbar 100 mg S

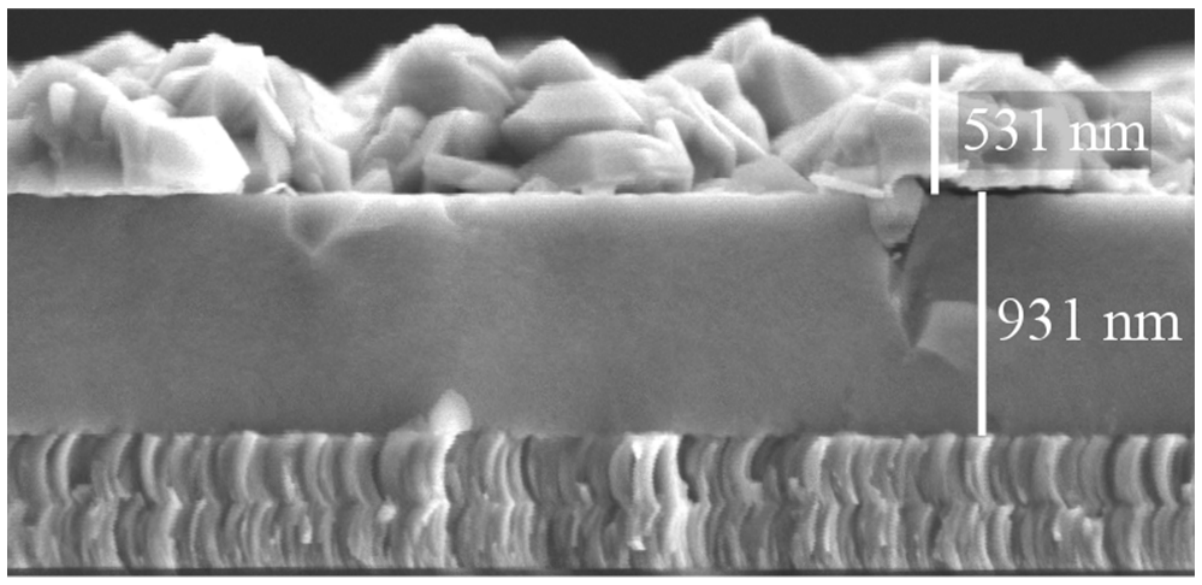

(d) 200 °C 30 min 1.5 mbar **500 mg S**

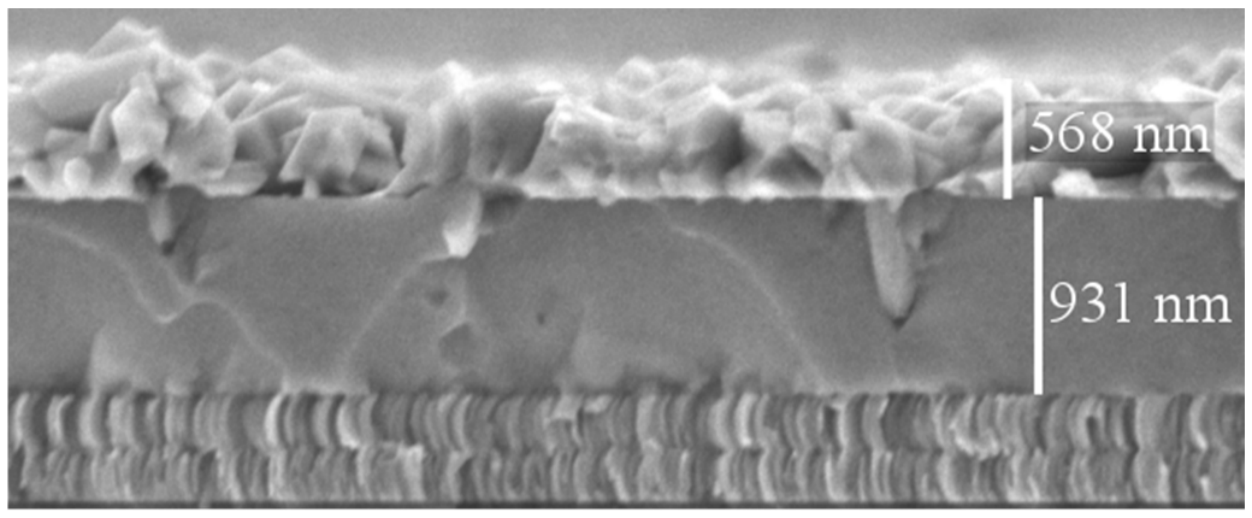

(e) **275 °C** 30 min 1.5 mbar **500 mg S**

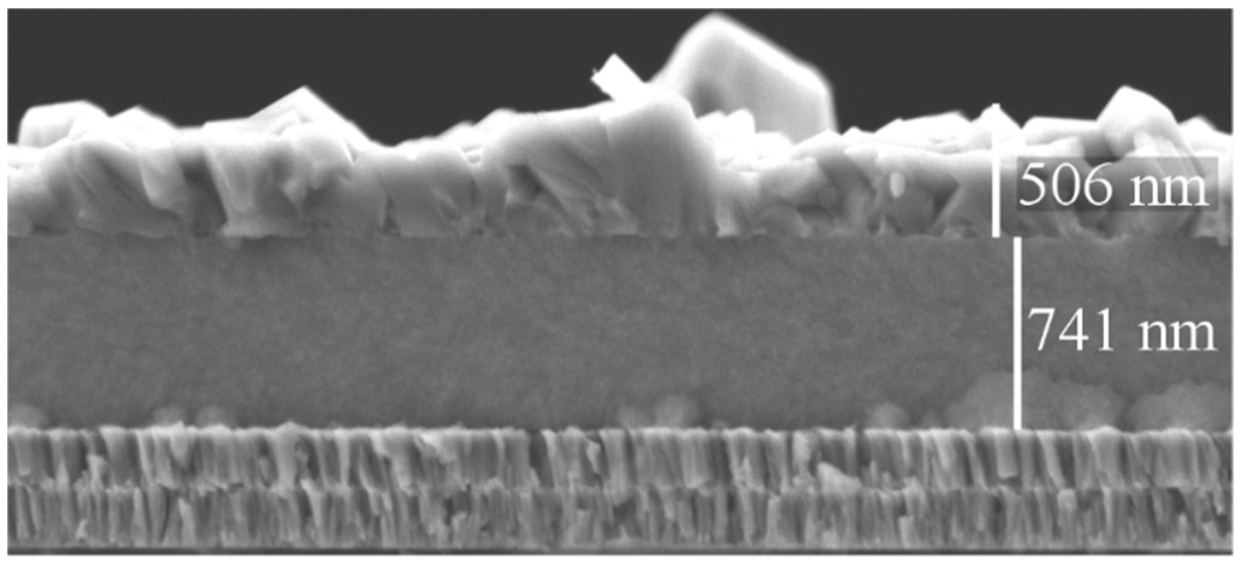

(f) **350 °C** 30 min **500 mbar 500 mg S**

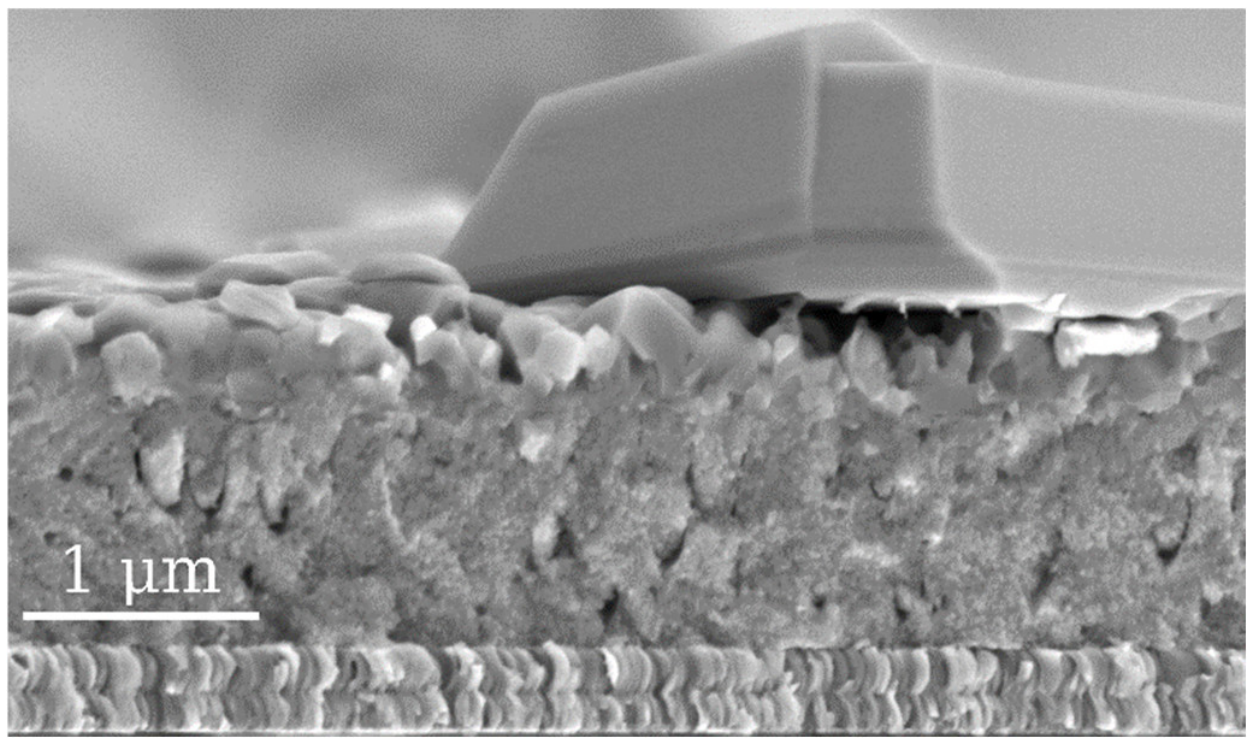

(g) **400 °C** 30 min **100 mbar** 100 mg S

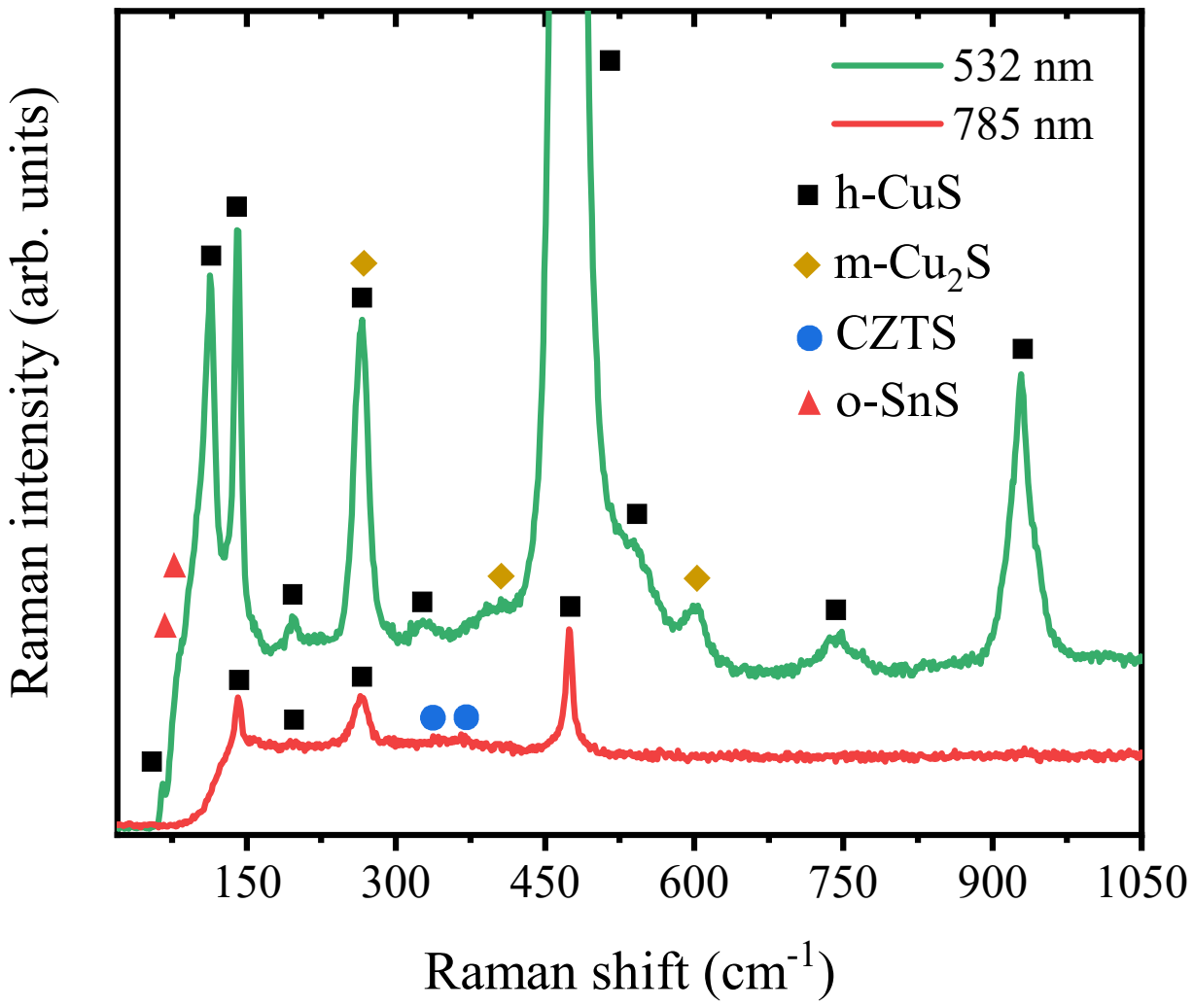

(h)

**Figure 2** – (a)-(g) Cross section SEM images showing the evolution of the double-layer morphology during the low temperature stages of a 2-stage annealing. The bold text highlights the changes to the reference conditions in (c); (h) Raman spectrum representative of sample (f).

Using Raman spectroscopy, the top layer was identified to be a mix of the hexagonal CuS phase (h-CuS) and the monoclinic $Cu_2S$ phase (m-$Cu_2S$),[61,62] with inclusions of the orthorhombic SnS (o-SnS) phase,[63] as shown in the Raman spectrum of **Figure 2 (h)**, obtained from the sample annealed at 350 °C in **Figure 2 (f)**. Even with an excitation wavelength of 785 nm, probing deeper into the samples, only a minor kesterite signal could be detected besides h-CuS at 350 °C,[64] indicating that the bottom layer has barely reacted below 350 °C. Further increasing the first stage temperature to 400 °C results in an intermixing between the two layers, as is visible in **Figure 2 (g)**. That intermixing continues into the second annealing stage (in our work mainly at 575 °C), resulting in a double-layered kesterite absorber, which can be seen in **Figure 3 (a)** in the next section. Besides the morphology, other features related to the double layer history during the annealing remain in the final absorber, as will be discussed later.

From the onset of the first stage, our findings reveal that an asymmetric crystallization profile arises during annealing. The upper layer is dominated by the formation of $Cu_{2-x}S$ phases, whereas the bottom layer appears to undergo a slower reaction. The two layers eventually mix partially at high temperatures, but the initial asymmetry does not vanish completely. In the early stages, as the temperature increases, and as soon as the available Cu is exposed to the sulfur in the annealing atmosphere, the h-CuS and m-$Cu_2S$ phases form. Given the high amount of available sulfur in the atmosphere during the first annealing stage, it is expected that the S-rich h-CuS phase will prevail over the m-$Cu_2S$ phase, which is in line with our Raman results of **Figure 2 (h)**. On the other hand, the h-CuS phase has the lowest melting point of the Cu-ZnS-SnS system, melting at a temperature of 507 °C[65] (for comparison, SnS melts at 882 °C, m-$Cu_2S$ melts at 1129 °C and ZnS melts at 1185 °C).[65,66] Therefore, the h-CuS may provide a viscous flow assisting the crystallization of the $Cu_{2-x}S$ phases, resulting in the large $Cu_{2-x}S$ crystallites visible near the surface. This is the dominant behavior seen in our system until 350 °C. At 400 °C, the temperature becomes sufficiently high and a visible reaction with the other binary sulfides occurs, as evidenced by **Figure 2 (g)**, and in agreement with earlier work.[66]

The findings above contrast with the dynamics of sputtered metal stacks, where the Sn melts very early at 231 °C, and thus the double layer arises through different dynamics, as thoroughly identified by Kim et al.[4] and discussed in the previous section. This emphasizes that the double layer formation appears to be a case-dependent phenomenon, despite its almost generalized appearance in kesterite solar cell research. It depends mostly on the precursor configuration and on the annealing conditions. However, a possible common aspect between this work and the referred sputtered metal stack works is that in these configurations the precursor is chalcogen deficient, so there is an inherent asymmetry during annealing, in that the chalcogen intake is heavily reliant on chalcogen gas arriving at the surface. In our work, the precursors generally have a ratio of S/metals = 0.65, whereas for sputtered metal stacks obviously (S,Se)/metals = 0. This inherent asymmetry could potentially be the main cause for the double layer dynamics. Similar to the present work, it has already been identified that the $Cu_{2-x}S$ phases promote a liquid-assisted growth from the surface at low temperatures (for an illustrative sketch of this effect, we refer the reader to ref. [21]). However, such a clear morphological asymmetry at different annealing stages from an initial uniform precursor stack of cosputtered Cu-SnS-ZnS has never been reported, to the best of our knowledge.

The different first-stage annealing parameters used in this work also give fundamental insights into the dynamics of the double layer formation. By increasing the amount of sulfur from 100 to 500 mg, as shown in **Figure 2 (b)**, we were able to show that the double layer is not caused by an insufficient amount of S during annealing – on the contrary, a higher S amount accentuated the double layer asymmetry. We have further confirmed this by increasing the S amount 16 times compared to the reference case of **Figure 2 (a)**, which is shown in the supporting information (SI) **Figure S1**. On the other hand, by increasing the pressure during the intermediate stage (from 1.5 mbar to 500 mbar in **Figure 2 (f)**), the chalcogen release rate is lowered due to the system pressure being higher than the sulfur vapor pressure, but the double layer asymmetry is still clearly visible. Therefore, the results appear to indicate that the double layer formation mechanism predominates at lower temperatures regardless of the first-stage annealing parameters. We note also that all the morphologies shown in **Figure 2** were obtained after a 30 min dwelling time at the designated temperature, followed by natural cooling, and are not obtained by quenching the samples as soon as the designated temperature is reached. While this approach has the disadvantage of not being the most accurate representation of the sample at each specific point during the complete annealing process, the main advantage is that it minimizes the nonuniformity that could occur by quenching due to the low thermal conductivity of the glass substrate (this effect has been discussed in detail in ref. [4]).

### *4.2 Characterization of double-layered absorbers from a two-stage annealing process*

After a complete annealing process and device fabrication, the finished morphology can be seen in **Figure 3 (a)**. This device, with the highest efficiency of 6.3%, was obtained using a 2S annealing process, with a first stage corresponding to **Figure 2 (c)**. The double layer interface in the kesterite is clearly visible, as indicated by the overlaid red line in **Figure 3 (a)**. The statistics on the performance parameters are given in **Figure 3 (b)**. The SIMS depth profile measurement of a sister sample of this device is shown in **Figure 3 (c)**. The SIMS results show a clear difference in the Cu, Zn and Sn signals between the top and the bottom layers, with a transition region which coincides with the double layer interface, as shown by the adjacent SEM image. Assuming that the matrix effects remain relatively unchanged across the depth, the results qualitatively indicate an increase in Zn concentration towards the back, and a correlated decrease in the Cu and Sn concentrations. Considering the low temperature formation studies shown in the previous section, the Cu depth profile is likely due to the strong accumulation of $Cu_{2-x}S$ phases occurring near the surface during the annealing process. Notably, we see also Na and O peaks aligning with the interface region of the double layer, further highlighting that this is indeed behaving as an interface in the kesterite (the acumulation of Na and O at the kesterite front and back interfaces has been well described in other works).[11] Furthermore, the SEM images reveal a high density of voids near the double layer interface of the absorber. In conjunction with the observations of **Figure 2**, this suggests that the double layer interface in the kesterite absorber follows a model similar to the Kirkendall effect.[67,68] The precursor matrix is sulfur deficient, and therefore the kesterite formation will rely on sulfur incorporation from the gas phase. On the other hand, the diffusivity of Cu is large compared to ZnS and SnS, leading to a high reactivity of metallic Cu with the S-rich atmosphere and a large Cu diffusion gradient and corresponding Cu migration towards the surface. Due to the higher Cu diffusivity, which is also higher than that of the S gas, the $Cu_{2-x}S$ growth will be driven by a continued Cu supply from the precursor bulk (rather than a S supply towards the bulk), in a self-regulating process that sets a sharp double layer interface at low temperatures.[69,70] This migration of Cu is known to result in an opposite flow of Cu vacancies towards the bulk.[70] During the second annealing stage at higher temperatures, there will be material exchanged through the double layer interface, and the interface position will be set by the difference in diffusivities of CuS, ZnS and SnS, in accordance with the Kirkendall effect, and also by the competing kesterite formation reactions. The vacancies resulting from this interdiffusion, (in particular the Cu vacancies, opposite to the Cu migration) may acumulate at the double layer interface, forming voids in this region. This phenomenon has been identified in several metallic alloys containing Cu, namely Cu-Zn and Cu-Ni,[67,68] and in the growth of $Cu_2S$ nanowires.[69] Specifically in CZTS, this model has been suggested by Platzer-Björkman et al. as a means to explain differences in void formation at the Mo back interface in cosputtering from metallic and from sulfide binary precursors.[71] To the best of our knowledge, we are unaware of any instances of this effect in $Cu(In,Ga)(S,Se)_2$ (CIGSSe) solar cells, despite the presence of metallic Cu. This could be related to the relatively wider stability region of CIGSSe compared to CZTSSe.

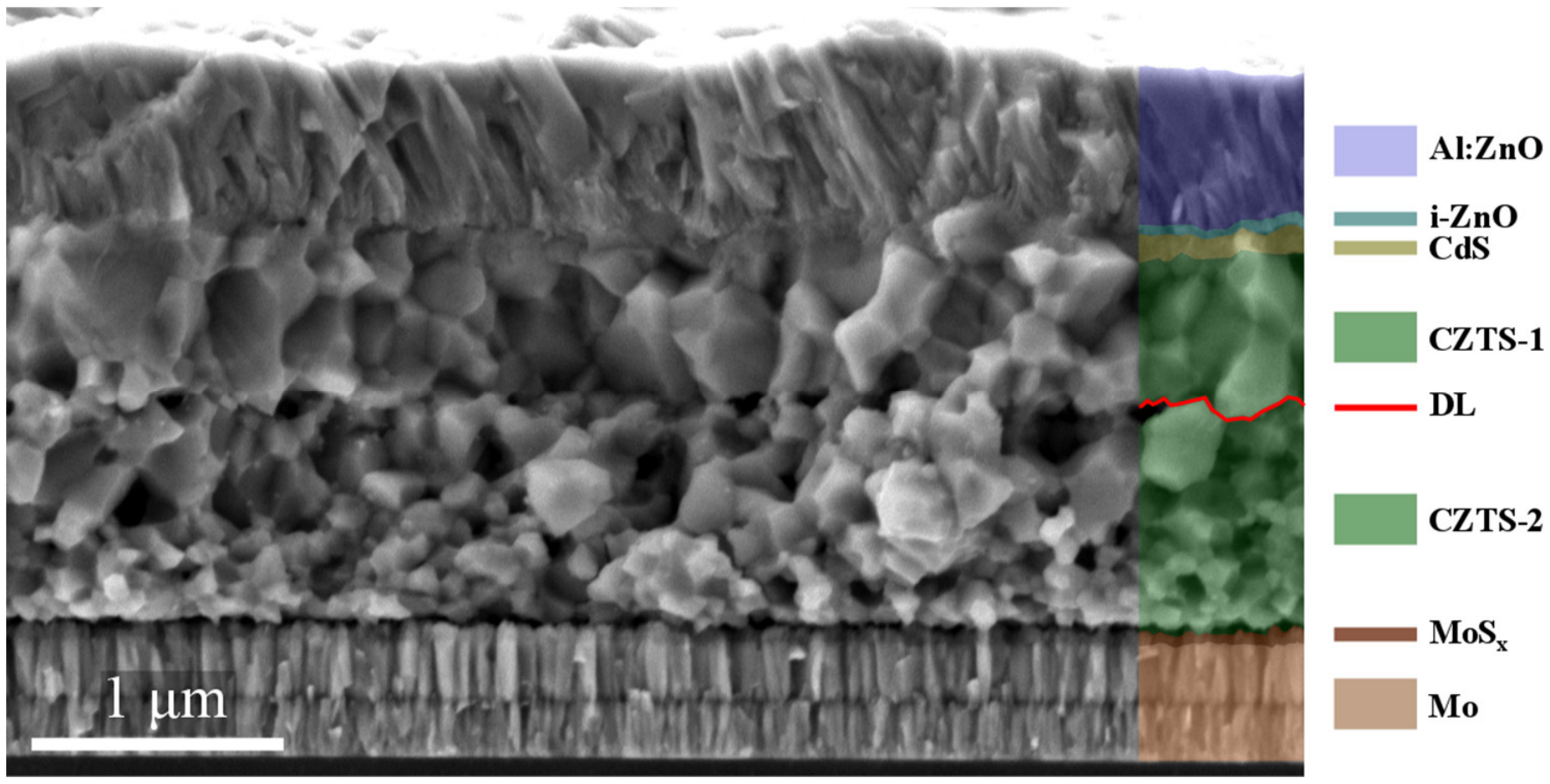

(a)

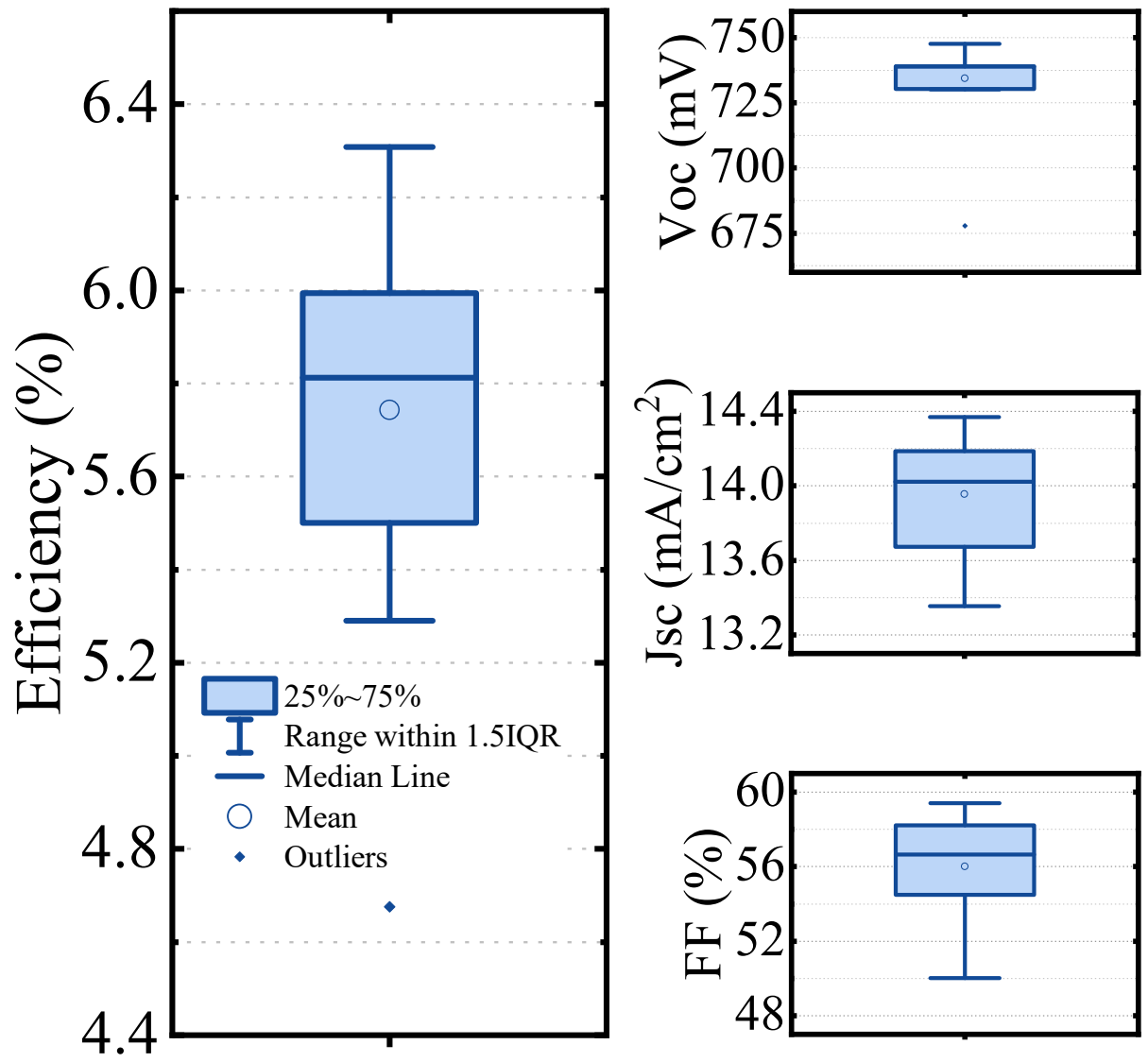

(b)

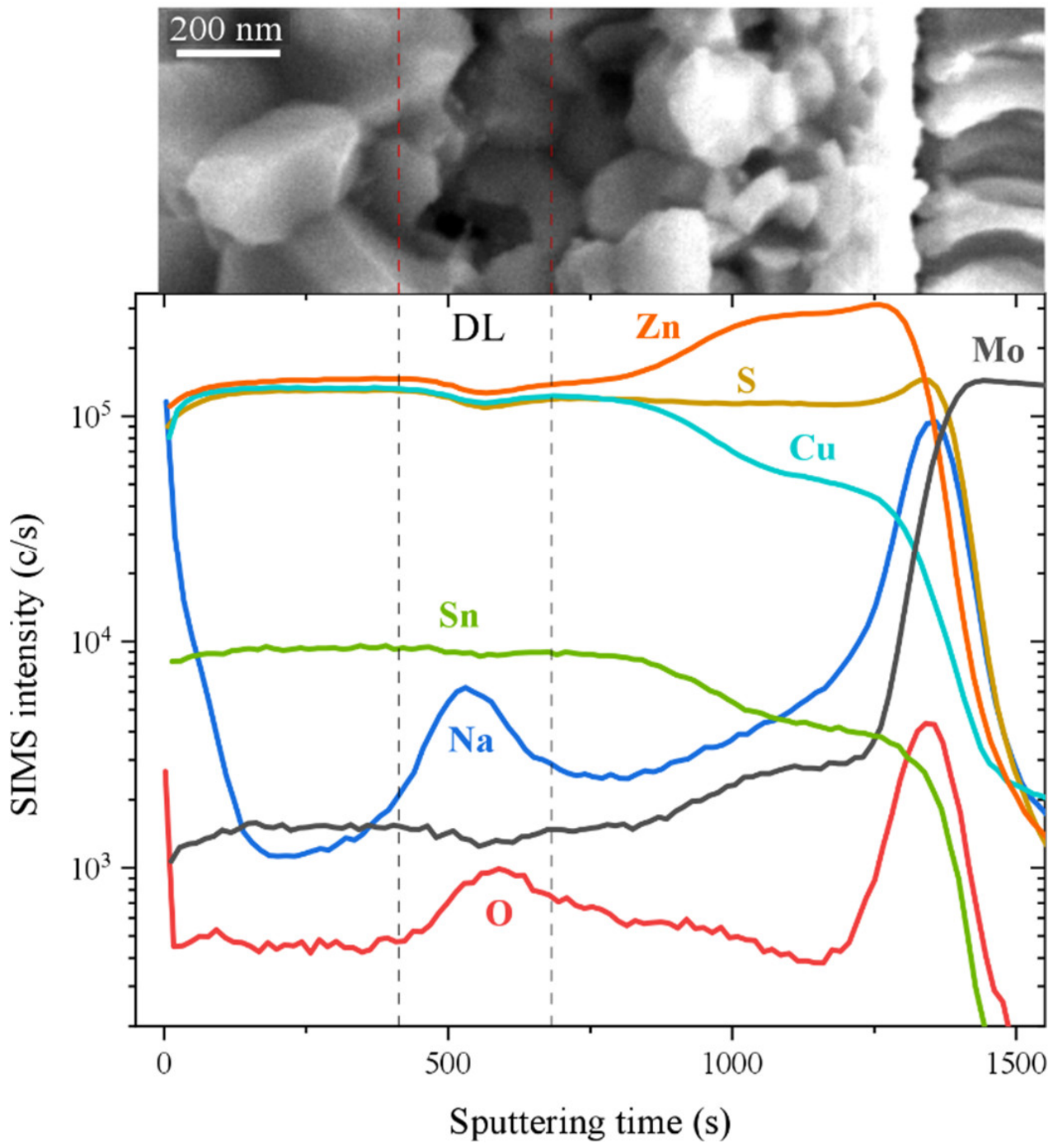

(c)

**Figure 3** – (a) SEM cross-section image of a double-layered kesterite device with 6.3% efficiency obtained from a two stage (2S) annealing process. The different layers are highlighted in color, including the double layer interface. (b) Corresponding boxplot statistics. (c) SIMS depth profile of an absorber layer similar to the one used in the device shown in (a). The double layer (DL) interface region is highlighted by the dashed line.

With the resulting composition deviations, confirmed by the SIMS results of **Figure 3 (c)**, the occurrence of secondary phases is likely, in particular near the back. To investigate this possibility, we have done multiwavelength Raman measurements on double-layered kesterite absorbers corresponding to cells reaching 6.6% efficiency.

Remarkably, despite the relatively high efficiency of these devices, we have confirmed the presence of traces of the secondary phases o-SnS and monoclinic $Cu_2SnS_3$ (CTS) (see **Figure S2**). We have also confirmed this using XRD, as shown later. Additionally, given the Zn-rich composition of this absorber, in particular towards the back, the presence of ZnS is also expected, even though this cannot be easily detected with XRD and without UV-Raman. This accumualtion of ZnS towards the back explains the correlated decrease in Cu and Sn signals in SIMS, which come predominantly from the kesterite matrix. We note, however, that the composition of the precursor corresponding to the absorber measured by SIMS is not excessively Zn-rich (we measured Cu/Sn = 1.86 and Zn/Sn = 1.10). These results indicate that the secondary phase occurrence is not only related to the composition region of the kesterite, but rather more likely also due to an incomplete reaction, especially for the bottom part of the kesterite. This is particularly evident in kesterite absorbers obtained with annealing temperatures lower than the nominal value used in this work (575 °C). At these lower temperatures, the double layer is more pronounced, and the grain size difference between the top and bottom layers is evident, as shown in **Figure S2**. Additionally, in **Figure S3** we provide a photograph of the absorber layers, the boxplot statistics and EQE measurement of the best cell.

### *4.3 Influence of annealing parameters*

To gain further insights on the double layer formation mechanism and its impact on device performance, we reoptimized the annealing step for a one-stage (1S) process in a different furnace, where the different annealing parameters could be tuned in a wider range. The differences between the 2S and 1S annealing processes can be found in **Table 2** and the **Experimental Section**. In **Figure 4**, we compare the morphology of the kesterite for different nitrogen pressures at the start of the annealing process. We note that the value of 555 mbar, corresponding to **Figure 4 (a)**, leads to an annealing at near atmospheric pressure due to gas expansion in the sealed quartz tube at 575 °C. By lowering the initial pressure from 555 to 350 and 175 mbar, the double layer features and the asymmetry between the two layers become increasingly evident, as shown in **Figure 4 (b)** and **(c)**. In conjunction with the findings of the previous sections, we hypothesize that this is connected to the release of sulfur gas during the ramping stage of the annealing process.

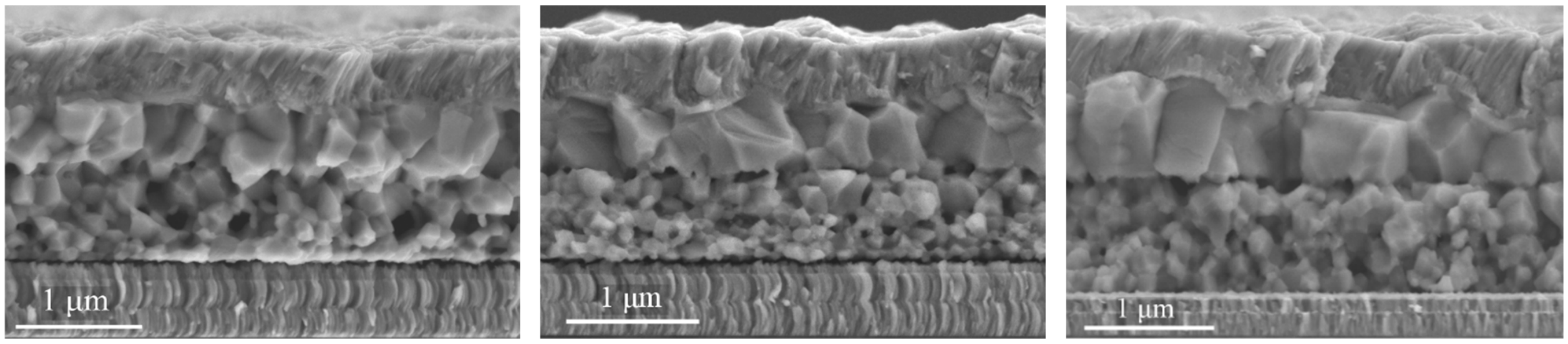

(a) (b) (c)

**Figure 4** – SEM images of CZTS devices showing the influence of the annealing pressure on the kesterite morphology, for initial $N_2$ pressures of (a) 555, (b) 350 and (c) 175 mbar. The corresponding best device efficiencies are (a) 2.3%, (b) 3.8% and (c) 5.7%.

As mentioned above, with increasing temperature the first new phases that form in our system are the $Cu_{2-x}S$ phases. As the reaction rate with the ZnS and SnS binaries is relatively limited at low temperatures, the growth of the $Cu_{2-x}S$ phases will be mostly mediated by the availability of sulfur in the atmosphere, which directly depends on the initial pressure in the tube. The lower the initial pressure, the higher the sulfur evaporation rate at low temperatures, leading to a higher double layer asymmetry in the early stages of the annealing process. As shown in the previous section, this asymmetry cannot be fully removed at the dwell temperature used (575 °C), which produces the morphology trend shown in **Figure 4**. An implication of this finding is that the sulfur availability directly controls the double layer dynamics during the ramping stage of the annealing. To confirm this, we have also performed the same annealing procedure without any S source, and indeed found no evidence of a double layer formation on the annealed sample. The morphology and corresponding Raman characterization of this sample are shown in **Figure S4 (a)** and **(b)**, respectively. Naturaly, due to the lack of sulfur, that sample consists of a mixture of kesterite and several secondary phases, and would not be very useful for device processing. However it points to the possibility of using a pre-annealing in an inert atmosphere, as discussed in **Section 2**. In our setup, this is not something we can practically do in a single annealing, but it could be an interesting possibility for future work. On the other hand, this finding also points to the possibility that the double layer can be related to the initial sulfur deficiency in the cosputtered precursor matrix, which leads to a dependence on the sulfur from the gas phase, and thereby to a large asymmetry in the sulfurization

Remarkably, we found a clear increase in device performance with the decrease in annealing pressure, with the sample annealed at low pressures showing the best performance of the series. The best efficiencies were 3.0% at 555 mbar, 3.8% at 350 mbar and 5.7% at 175 mbar, as shown by the device statistics in **Figure S5**. Moreover, we systematically found the low pressure annealing to yield higher efficiencies and more reproducible results even for a wide range of starting (precursor) compositions, as detailed by the device statistics in **Figure S6.** This apparently surprising result highlights the complex interplay of the double layer dynamics, with the simultaneous occurrence of compositional and growth asymmetries in the samples during annealing. We note that these results do not necessarily mean that in general a low-pressure annealing is better suited for fabricating kesterite solar cells, and could in principle constitute only a local maximum in our optimization space. However, the results do show that the double layer dynamics can significantly impact the performance of the resulting devices, given the clear link between the annealing pressure, the double layer formation and the corresponding device statistics of **Figure S5** and **Figure S6**.

Furthermore, we have fundamentally found that within a reasonable parameter space, the double layer formation is nearly ubiquitous in our cosputtering process. For instance, in conjunction with the device statistics of **Figure S6**, we have also observed that the double layer occurs regardless of the kesterite composition. In **Figure S7** we compare the double layer morphologies of two samples from a slightly Cu-rich precursor (Cu/Sn of 2.10) and from a Cu-poor precursor (Cu/Sn of 1.69). Although the morphologies are visibly different, both clearly show double layer features. The sample with higher Cu content shows a higher asymmetry with visibly larger grains in the upper layer. This again matches well with our low temperature formation model proposed above – the higher Cu content leads to a more pronounced formation of $Cu_{2-x}S$ phases at the surface, thereby creating a higher double layer asymmetry. In all these experiments, a device-relevant Zn-rich composition was used, resulting in a ZnS secondary phase segregation at the bottom of the films. Given that presence of ZnS, one could wonder if the double layer occurrence could be caused by the incomplete reaction of this ZnS phase. However, we still see a double layer segregation in Zn-poor compositional regions, where other secondary phases are segregated, as shown in **Figure S8**. In general, we found that near the compositional range that allows high-efficiency devices, the double layer always occurs in our cosputtering approach. As for the annealing temperature and dwelling time, we also systematically found them to have very little utility in resolving the double layer. By increasing the annealing temperature from 575 °C to 595 °C, we were able to obtain single-layered absorbers with large grains, as shown in **Figure S9**. However, despite the apparent improvement in morphology, the corresponding devices always performed significantly worse, with efficiencies below 2%. A similar trend in morphology and performance degradation was found with an increase in dwelling time (not shown here). One possible explanation is that this could be revealing a fundamental trade-off between the dynamics of the double layer and the final kesterite defect composition. As mentioned above, our kesterite absorbers exhibit several different secondary phases, which are not limited to the specific compositional region of the kesterite, in particular in the bottom layer, suggesting an incomplete reaction. Therefore, relatively higher temperatures and dwelling times would be favored. However, further increasing these parameters likely leads to a drop in sulfur partial pressures and kesterite decomposition, and consequently to the appearance of detrimental defects, as discussed in ref. [15]. In particular, we note that the most recent results for the best-performing sulfide CZTS devices were obtained using temperatures lower than 580 °C and annealing times below 15 min.[27,30] Therefore, we estimate that tuning the annealing temperature and dwelling time should be insufficient to resolve the double layer issue. One other possible way of dealing with this trade-off would be to increase the heating rate in our 1S annealing setup. By increasing the heating rate, the effective time within the $Cu_{2-x}S$ formation region is lower, and therefore the double layer formation should be minimized. In **Figure S10**, we compare the morphology of the kesterite using our reference heating rate of 20 °C/min and a heating rate of 40 °C/min, and in **Figure S11** we present device statistics for a series comparing the heating rates of 20, 30 and 40 °C/min for the same base precursor. As shown in **Figure S10**, the double layer appears to visibly improve, but it is not completely removed. An interesting aspect of this series was the systematic improvement in $V_{oc}$ with increased heating rate, up to above 600 mV, as shown in **Figure S11**. However, the device performances were comparable, and so far we were not able to demonstrate improvements using this approach. Therefore, we suggest that tuning the heating rate at low temperatures could be an interesting option for future work.

### *4.4 Influence of absorber thickness*

In this section we show that the absorber thickness has a crucial impact on the double-layer dynamics and device performance. In **Figure 5**, a series comparing the kesterite morphology for different absorber thicknesses is shown. The absorber thickness was changed by varying the cosputtering time from 10 min (300 ± 50 nm) to 38 min (1400 ± 200 nm), in several steps. We find that the double layer is not present for very thin samples (**Figure 5 (a)-(c)**, corresponding to a critical thickness below 600 nm), and a transition point exists around 600-750 nm (**Figure 5 (d)**) where the double layer interface appears in the kesterite. For thicker samples above 750 nm, we found that the double layer always occurs within our 1S reference annealing.

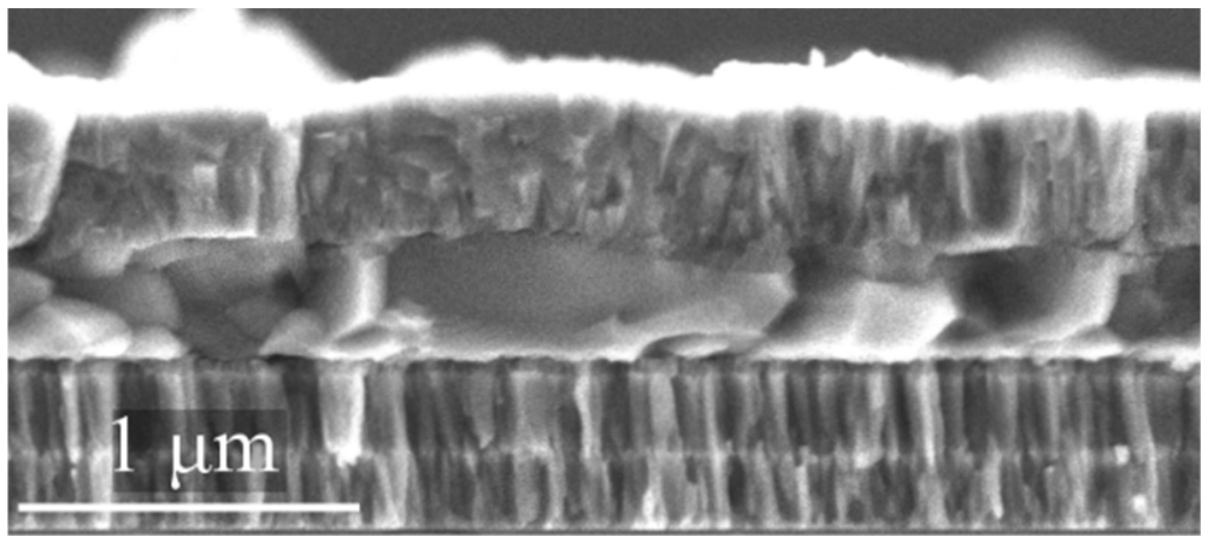

(a) 10 min (300 ± 50 nm)

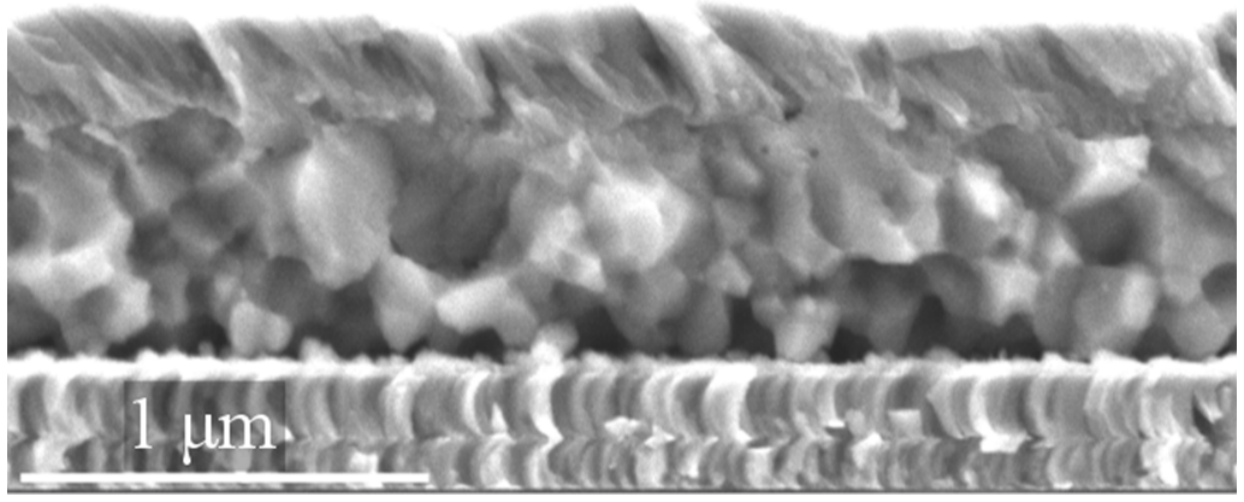

(b) 12 min (475 ± 20 nm)

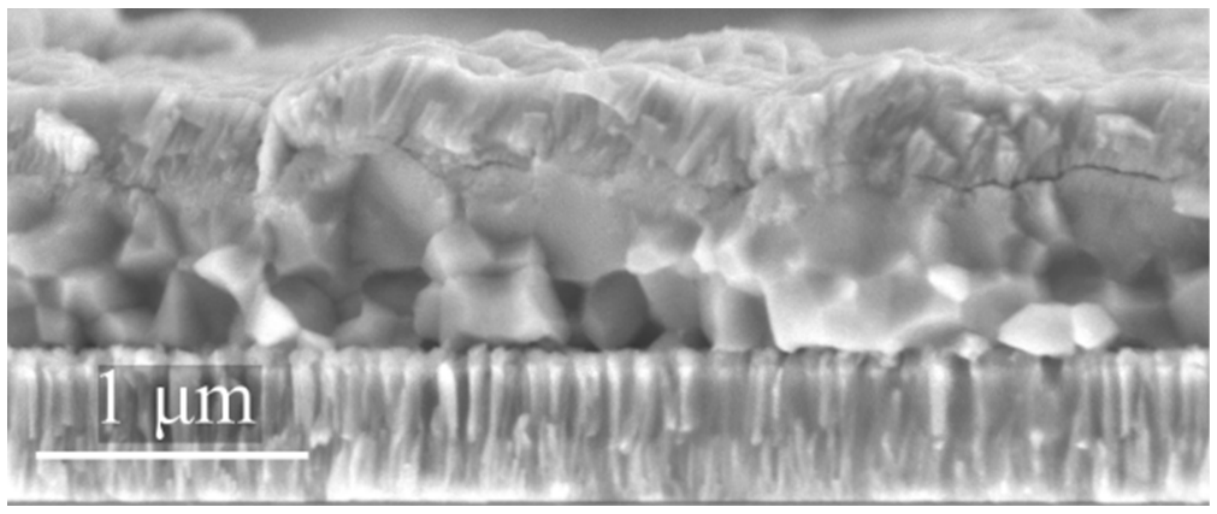

(c) 15 min (550 ± 20 nm)

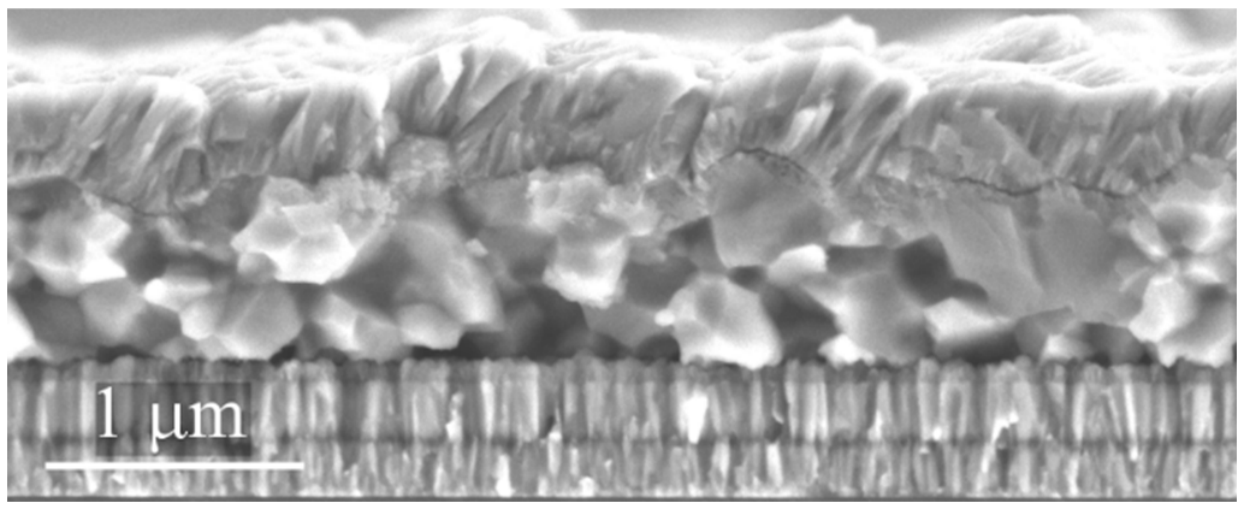

(d) 17.5 min (625 ± 90 nm)

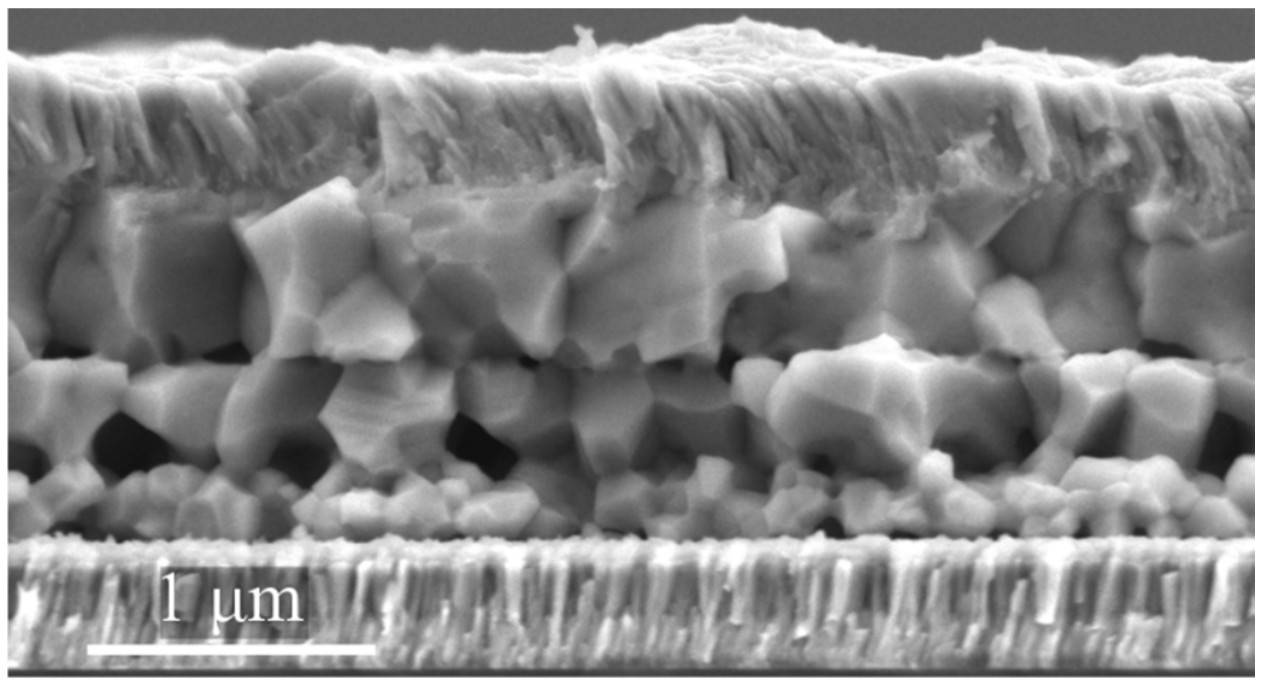

(e) 28 min (1100 ± 100 nm)

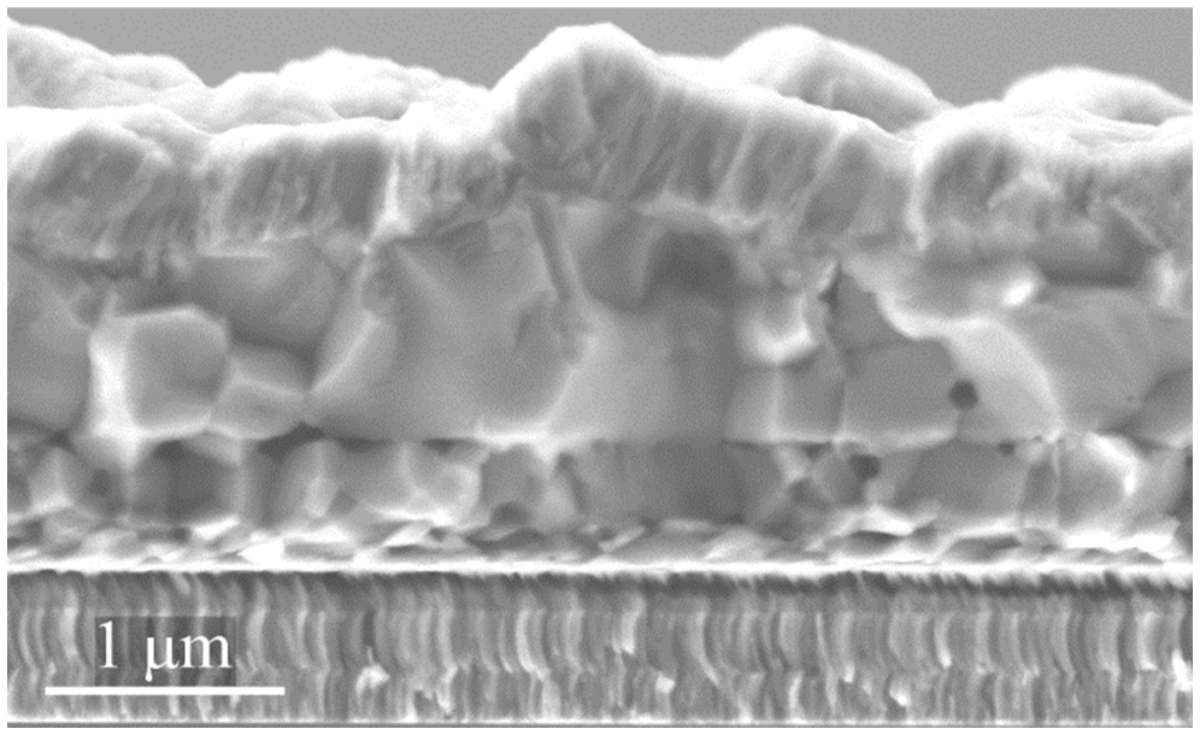

(f) 38 min (1400 ± 200 nm)

**Figure 5** – SEM cross-section images of devices comparing the kesterite morphology for different cosputtering times of (a) 10 min (b) 12 min, (c) 15 min, (d) 17.5 min, (e) 28 min and (f) 38 min. The corresponding thicknesses are indicated in parentheses.

To complement the results of **Figure 5**, we have also investigated the double layer formation in thin samples for different annealing conditions. In **Figure S12** we show that progressively lowering the annealing pressure from 175 mbar to 100 mbar will again lead to a sharply defined double layer in the kesterite, in particular in a 2S annealing. Therefore, despite the apparent improvements, we can establish that the overall double-layer formation dynamics are still occurring in thin samples. Based on our results, the likely explanation for the thickness dependence of the double layer is again based on the formation of the surface $Cu_{2-x}S$ layer. The formation of this $Cu_{2-x}S$ layer is governed by the migration of the available Cu to the surface and reaction with the S available in the atmosphere. By preventing the direct reaction of the S gas with the bottom part of the precursor, the $Cu_{2-x}S$ layer effectively acts as a sulfur diffusion barrier. The thicker the $Cu_{2-x}S$ layer, the sharper the resulting double layer asymmetry. The thickness of the $Cu_{2-x}S$ layer is determined by the Cu and S availability, and by the low-temperature annealing conditions, in a self-regulating process. In thick samples (>1000 nm), for our reference 1S and 2S conditions, and for device-relevant compositions (near Zn-rich and Cu-poor), the upper layer/bottom layer thickness proportion ranges from a minimum of 33%/67%, respectively, up to a maximum of 50%/50%. By reducing the overall thickness, the $Cu_{2-x}S$ thickness is reduced in a similar proportion, leading to a weaker blocking of the reaction of S with the precursor. For very thin samples, this $Cu_{2-x}S$ becomes essentially nonlimiting, leading to a uniform kesterite layer. However, even in thinner samples the background problem of S deficiency in the precursors still remains, which explains the persistence of the double layer as we have shown in **Figure S12**.

To determine the effects of thickness and the impact of the double layer on the performance of the devices, we prepared a series of devices from precursors fabricated and processed in the same batch, but with different absorber thicknesses. The precursors and resulting devices are otherwise identical within the limits of our processing line. The device statistics of the series are shown in **Figure 6 (a)**, and the statistics for the equivalent one-diode electrical circuit parameters are shown in **Figure 6 (b)**. The diode parameters were estimated under illumination using Sites' method [72]. Furthermore, **Figure 6 (c)** displays the corresponding thickness calibration as a function of cosputtering time. The best cell results of the series are summarized in **Table 3**. Additionally, the IV curves of the best cells and the fitting of the data for the one-diode model are compiled in **Figure S13**. Our solar cell performance data reveal two clear trends: the progressive decrease in $J_{sc}$ and increase in FF with decreasing thickness of the absorber layer. Besides the double layer morphology, there are multiple other factors at play when varying the absorber thickness. First, the decrease in $J_{sc}$ is expected as thicknesses below 400 nm are clearly insufficient to absorb all the incoming photons. Assuming an absorption coefficient of $10^4$ $cm^{-1}$ for CZTS, only 26% of the incoming photons above the bandgap would be absorbed for a thickness of 300 nm, and a thickness of at least 2000 nm would be needed to reach 90% absorption. However, in our case, the numbers are slightly more favorable, as the samples with a thickness of 300 nm (10 min cosputtering) reach above 15 mA/cm$^2$, which is already around 50% of the Shockley-Queisser $J_{sc}$ limit. Therefore, a decrease in $J_{sc}$ due to optical losses cannot be the only explanation. Given that our $MoS_2$ layer is very thin, one factor could be that some photon recycling can occur by reflection at the Mo back interface. However, the typical reflectivity of Mo in air is below 0.6, and for CIGS solar cells it has been estimated that a $J_{sc}$ loss of up to 6 and 8 mA/cm$^2$ occurs due to absorption (poor reflectivity) of the Mo back contact, for CIGS thicknesses of 500 and 300 nm, respectively.[73] Put another way, for CIGS thicknesses of 500 and 300 nm, and considering a reflectivity of 0.5 for Mo, only a gain in $J_{sc}$ of about 1.7 and 3 mA/cm$^2$ is expected to occur, respectively, compared to neglecting the back-contact reflectivity.[74] For that reason, we expect the back contact reflectivity to be only a small contribution to our results for thin samples. We must therefore look for clues in the other device parameters.

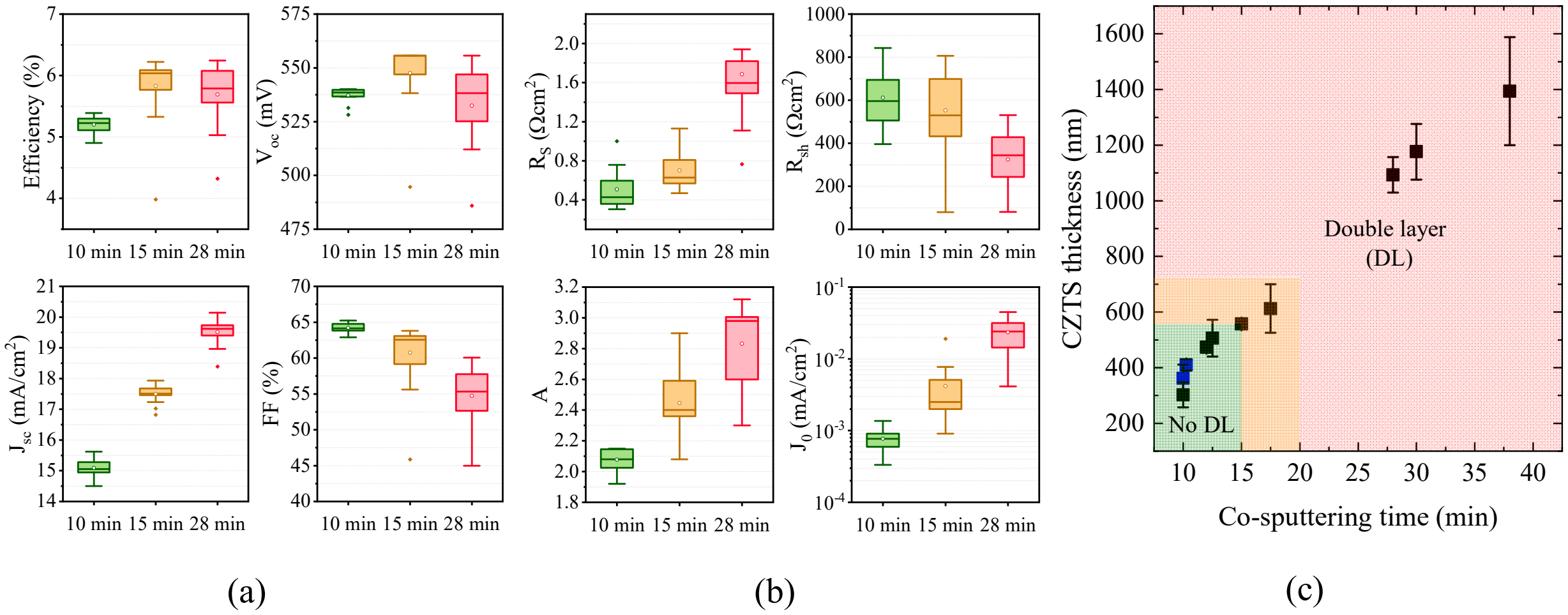

**Figure 6** – (a) Boxplot showing the device performance statistics for different absorber thicknesses; (b) One-diode model fitting parameters determined from the light IV curves using Sites' method [72]; (c) Corresponding thickness calibration plot, showing the CZTS absorber thickness as a function of cosputtering time. The blue points were obtained from CZTS grown on monocrystalline Si substrates.

**Table 3** – Solar cell parameters for the best cells of the thickness series of **Figure 6**.

| Sample | Eff (%) | $V_{oc}$ (mV) | $J_{sc}$ (mA/cm$^2$) | FF (%) | $R_S$ (Ωcm$^2$) | $R_{sh}$ (Ωcm$^2$) | $A$ | $J_0$ (mA/cm$^2$) |
|---|---|---|---|---|---|---|---|---|
| 10 min (300 nm) | 5.39 | 538 | 15.6 | 64.1 | 0.42 | 504 | 2.03 | $5.7\times10^{-4}$ |
| 15 min (550 nm) | 6.22 | 556 | 17.8 | 63.0 | 0.47 | 732 | 2.41 | $2.5\times10^{-3}$ |
| 28 min (1100 nm) | 6.25 | 556 | 20.1 | 55.8 | 1.61 | 404 | 2.94 | $1.8\times10^{-2}$ |

The FF trend is also clearly visible in **Figure 6 (a)**, suggesting a degradation of the electrical characteristics of the devices with increasing absorber thickness. This trend matches the increase in the series resistance $R_S$ and decrease in shunt resistance $R_{sh}$ under illumination, as shown in **Figure 6 (b)**. Notably, a decrease in shunt resistance for very thin devices is commonly reported, due to the potential occurrence of pinholes, higher roughness-to-thickness ratio, or smaller absorber grain size.[74,75] Here, in the absence of such morphological defects, we observe the opposite trend, i.e., $R_{sh}$ increases for the thinner devices. On the other hand, the $V_{oc}$ does not show a clear trend. To further resolve the results of **Figure 6 (a)**, we have compiled a large statistical dataset of $V_{oc}$ and FF values corresponding to 678 individual cells across 43 samples fabricated in our group over the course of approximately one year. The results are shown in **Figure 7**. The selection criteria for these samples was that they were processed using a reference 1S or 2S annealing, and at least one cell per sample reached an efficiency of 5%, except for thicknesses below 500 nm, where a 4% cutoff was used. We then categorized the results by the thickness of the absorber layer, as shown in the abcissa of **Figure 7**. Analyzing the median lines confirms that the FF decreases for thicker samples (above 1000 nm), by at least 5 percentage points. On the other hand, for samples thinner than 500 nm and in the range 550-750 nm, the difference does not appear to be significant. As for the $V_{oc}$, we find no difference between the three thickness categories.

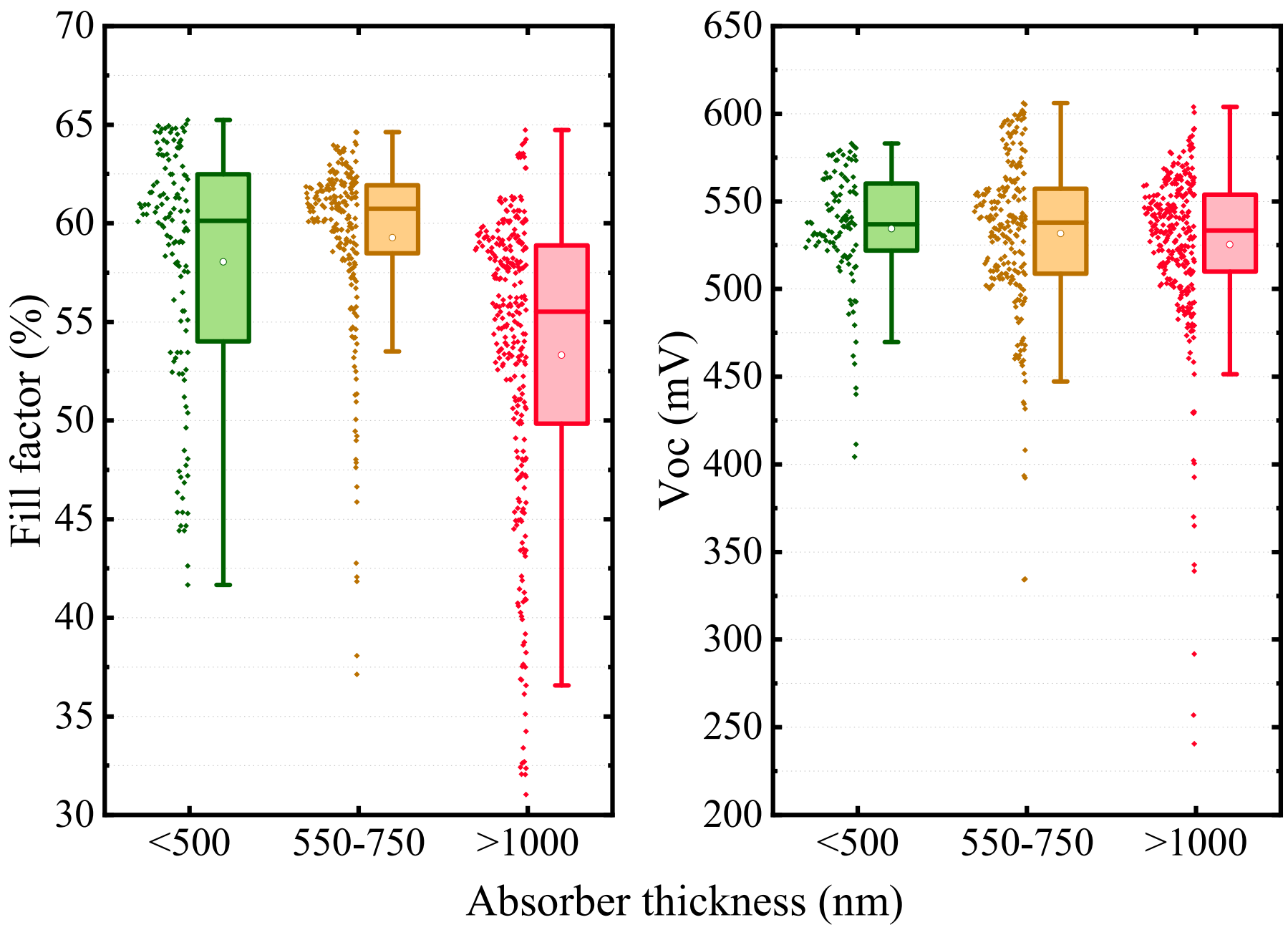

**Figure 7** – Statistical distribution of FF and $V_{oc}$ categorized by absorber thickness. The dataset consists of 678 cells with an area of 0.09 $cm^2$ from samples annealed using the reference 1S and 2S processes. The samples were selected based on a minimum efficiency of 5% (4% for thicknesses below 500 nm).

However, this flat trend in $V_{oc}$ is unexpected. In general, the $V_{oc}$ of any solar cell will decrease as the thickness tends to zero, unless surface recombination at the front and back surfaces is completely suppressed (in which case the $V_{oc}$ slightly increases with decreasing thickness).[76] This is simply because the generated minority carrier density (in this case electron density) will be higher near the back contact when compared to a thicker absorber, which leads to a higher surface recombination rate if the surface recombination velocity is not negligible. In turn, a higher surface recombination rate will decrease both the $V_{oc}$ and the $J_{sc}$. In fact, it is well known in the CIGS field that for ultrathin devices (<500 nm), the main performance limitations are optical losses due to thinning and back contact recombination due to the non-ideal Mo/CIGS interface, which result in lower $V_{oc}$ and $J_{sc}$. In CIGS, these limitations can be mostly overcome by using a back surface field based on a Ga/(Ga+In) grading or by using back contact passivation strategies.[73–75] However, in kesterite solar cells, a back grading is not trivially available, and the kesterite/Mo interface is known to be more unstable than its CIGS counterpart.[77] Therefore, in our case, we do not expect our back surface recombination velocity to be better than in CIGS devices. The combination of these factors should explain the decreasing $J_{sc}$. However, they would also leave us with the expectation that the $V_{oc}$ should decrease, which is contrary to our observations – the $V_{oc}$ does not change. We additionally note that this $V_{oc}$ degradation with decreasing thickness has already been reported for single-layered CZTS devices by Ren et al..[78] Therefore, in our case there must be another factor contributing to an increase in the $V_{oc}$ for lower thicknesses, which counterbalances the aforementioned $V_{oc}$ degradation and results in the flat trend of **Figure 6 (a)** and **Figure 7**. Given our morphology observations, this improvement could come from the double layer dynamics, namely with the disappearance of the double layer interface in thinner samples. As mentioned above, the improvement in shunt resistance $R_{sh}$ in thinner samples will lead to an increase in $V_{oc}$. Furthermore, analyzing the ideality factor $A$ and reverse saturation current densities $J_0$ in **Figure 6 (b)**, we see that $A$ and $J_0$ increase for thicker samples, which indicates an increased nonradiative recombination in the depletion region for thicker samples. The $A$ value obtained for our thin samples (300 nm) of 2.03 is typical of state of the art CZTSSe devices (where $A$ is generally below 2), whereas in other thin film technologies an $A$ as low as 1.3 has been demonstrated.[27] Applying the same fitting in the respective dark curves for all thicknesses (not shown here for briefness), we obtain similar $R_S$ but the $R_{sh}$ is 1 order of magnitude higher, the $A$ is always below 1.9 and $J_0$ is at least 2 orders of magnitude lower, indicating that a voltage dependent shunting is not occurring, but instead there is a voltage dependence on the photocurrent collection efficiency.[72] From this data alone, the exact recombination mechanism cannot be resolved, as there can be contributions from multiple sources such as bulk defects, tunneling-assisted recombination and the CZTS/CdS band alignment.[72] Although associating this degradation with the double layer seems plausible, considering the multiple factors at play and their complexity we must note that to fully validate this claim we would have to perform additional characterization that would allow us to specifically probe the dominant recombination mechanisms and the optoelectronic properties of the double layer interface and of the top and bottom layers, which is not trivial, and would be beyond the aim of this work. A detailed material characterization study, comparing thin kesterite absorbers and thick (double-layered) absorbers is therefore suggested as future work. Interestingly, the thicker samples, which exhibit the double layer, show $J_{sc}$ values up to 20 mA/cm$^2$ (in **Figure 6 (a)**), which is comparable to record sulfide devices.[27,30] Therefore, we are able to infer that the bottom layer should still be contributing to the overall current output of the device, in spite of the different morphology, composition and presence of secondary phases. As for the FF, similar observations can now be made coherently with the $V_{oc}$ and $J_{sc}$ results. The FF is expected to decrease with decreasing thickness as it follows the $V_{oc}$,[79] however, for very thin devices (<300 nm), the depletion width may extend throughout the whole absorber layer, which increases the collection efficiency and the FF.[74,79] This can explain the slight improvement in FF for the 10 min (300 nm) sample shown in **Figure 6 (a)**, although, as shown by the large statistics of **Figure 7**, this effect should at best be relatively small. For thicker samples, however, the clear degradation in FF is again unexpected, considering that the $V_{oc}$ is flat. This would again be consistent with the effects of the double layer, in particular due to the effects of the double layer interface but also due to the small grain size and accumulation of secondary phases at the back.

Regarding the data of **Figure 7**, we should also note that, given the long processing timeline of the dataset (approximately one year), there can be some variability between the samples in terms of the kesterite synthesis and composition, the CdS chemical bath deposition and the transparent conductive oxide (TCO) conditions. We assume that such variability occurs spontaneously and is randomly distributed across our dataset. Moreover, we note that using the median line should be a more robust metric to analyze the central trends of our dataset, as the mean value can get severely skewed by outliers.

### *4.5 Effects of postdeposition heat treatments in double-layered absorbers*

It is well-known in the kesterite field that a postdeposition heat treatment (or post-annealing, PA) can change the properties of the kesterite and the device interfaces, which may lead to improvements of the device performance. Therefore, it is illustrative to complement the results of the previous sections with additional device statistics for post-annealed devices. There have been multiple different PA strategies reported for CZTS, CZTSe and CZTSSe kesterites, and discussing it is beyond the scope of this work. In our group, we have optimized two types: one PA at 150 °C in air for 15 min after the CdS deposition, and another PA at 275 °C in $N_2$ for 1 min, either after CdS or in a full cell (with similar results). A detailed characterization of these post-annealings will be reported in a future publication. In **Figure 8**, the results of these post-annealing treatments are shown for a device with a thickness around 550 nm (15 min cosputtering). Interestingly, the improvement comes from different parameters in each case.

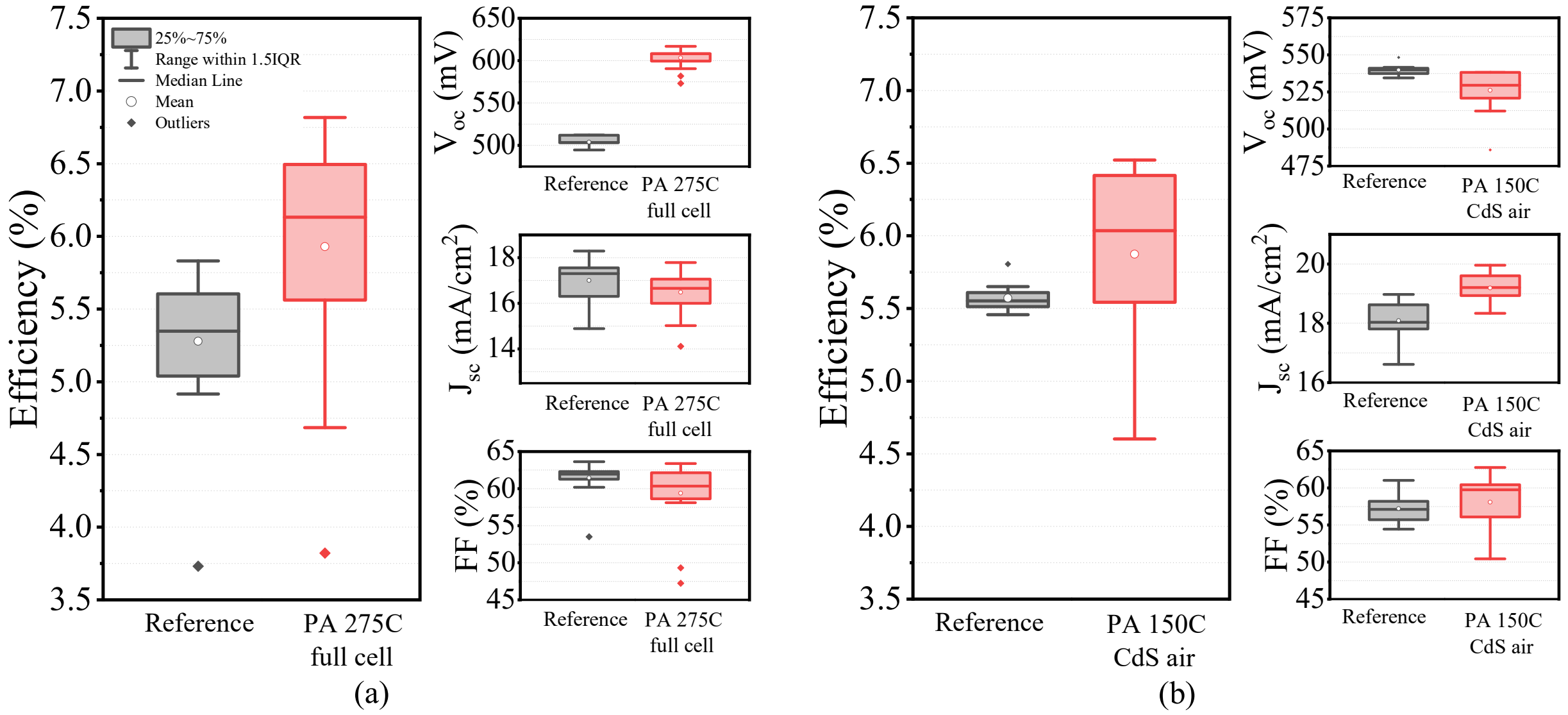

**Figure 8** – Boxplot statistics showing the effects of post-annealing (PA) treatments in devices with a 550 nm-thick kesterite (15 min cosputtering), processed using a 1S annealing. The conditions are (a) PA of the full cell at 275 °C for 1 min followed by quenching and (b) PA after CdS deposition at 150 °C for 15 min followed by quenching.

For the full cell PA in $N_2$ in **Figure 8 (a)**, there is a clear 100 mV improvement in the $V_{oc}$. The conditions of this PA are very similar to the work by Yan et al..[27] In that work, it has been discovered that for these PA conditions, the $V_{oc}$ improvement is related to a reduction in heterojunction interface recombination. Therefore, it is plausible to assume that at least a part of the improvement in the $V_{oc}$ in **Figure 8 (a)** comes from a reduction in the CZTS/CdS interface recombination. Therefore, with the results of **Figure 8 (a)** we are able to indirectly prove that our devices are at least partially affected by interface recombination. The importance of this result is that it further validates the discussion on the thickness variation in the previous section – if our devices (without any PA) are limited by interface recombination, then a reduction of $V_{oc}$ with thickness should be expected. As mentioned above, this was not the case, suggesting that the double layer improvements for thinner samples could be contributing to a counterbalancing increase in the $V_{oc}$. On the other hand, **Figure 8 (b)** reveals that the improvements for the low temperature PA in air after CdS come from improvements in $J_{sc}$ and FF. In particular, the $J_{sc}$ reaches up to 20 mA/cm$^2$, which is the same level of the thick sample shown in **Figure 6 (a)** despite the kesterite absorber being half the thickness. Thus, this PA experiment highlights that our thick samples have a poor collection efficiency across the bulk, again suggesting a possible negative influence of the double layer. A sister graph to **Figure 8** but with thick samples (>1000 nm) can be found in **Figure S14**, and in **Table S1** we compile the one-diode model parameters for the 8 best cells corresponding to each case. We do not see a clear trend regarding the effects of the PA for different thicknesses, and the overall parameter comparison is very similar to the results of **Figure 6** and **Figure 7**, meaning that, on average, thin post-annealed samples also exhibit higher FF and similar $V_{oc}$ compared to thick post-annealed samples. This suggests that despite the improvements, there is a differentiating background aspect that is not being changed by the post-annealing treatments. This aspect could be the double layer, as these low temperature heat treatments will not cause a major recrystallization of the kesterite, and are likely impacting mostly the interfaces and grain boundaries. Still, there is a possibility that the PA treatments can change the properties of the double layer interface, but this is not clear from our data.

Given the similar processing conditions of the series in **Figure 6**, we are basing our analysis on the assumption that the only difference between the samples is their thickness and the occurrence of the double layer. However, this does not necessarily mean similar values of the optoelectronic properties of the kesterite (namely bulk lifetime, carrier concentration, depletion width), precisely because of the complex dynamics of the double layer formation – with composition gradients, grain boundary changes and secondary phases. In any case, this is essentially still the general conclusion of this work: the possibility that the double layer formation is negatively impacting the performance of kesterite devices. To accurately decouple all the possible effects resulting from the double layer dynamics and their impact of the kesterite devices, we suggest that a detailed materials characterization study is still required as future work.

### *4.6 Influence of alkali metal availability (growth on Si substrates)*

Finally, it is fundamental in this discussion to study the inclusion of Na in the double layer dynamics, given its role as a fluxing (or surfactant) agent during the kesterite annealing process, through diffusion from the SLG substrate. To specifically isolate the effects of Na inclusion in the kesterite, we have constructed two case studies: one using 10 nm evaporated NaF on reference kesterite devices, and one using monocrystalline Si substrates, where Na is not present. A comparison of the resulting morphology of the kesterite is shown in **Figure 9**.

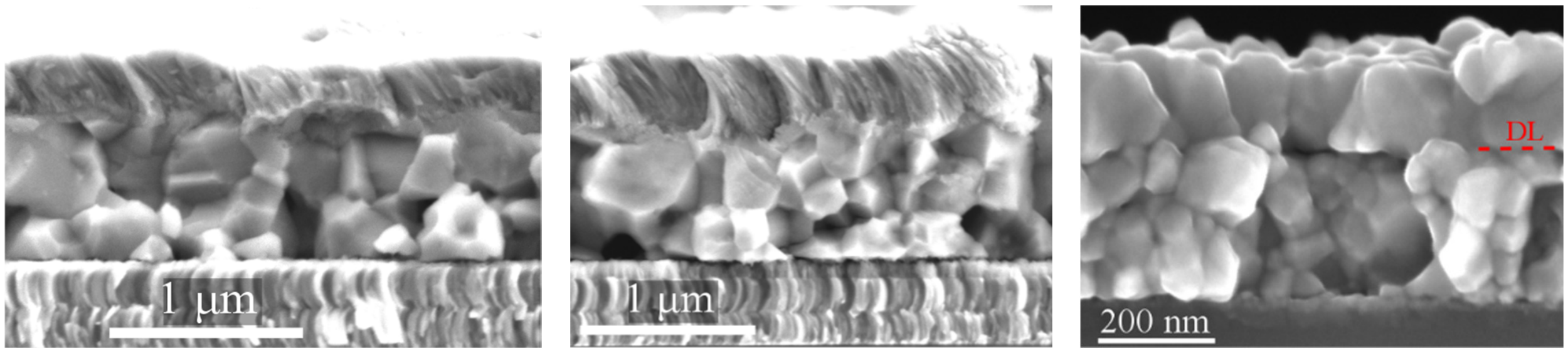

(a) No NaF (b) 10 nm NaF (c) CZTS on TiN/Si

**Figure 9** – SEM cross section images for standard kesterite devices with (a) no evaporated NaF (reference), (b) 10 nm evaporated NaF and (c) CZTS grown on a monocrystalline silicon substrate, without any Na supply. The double-layer (DL) interface is marked with the dashed red line. The corresponding best device efficiencies were (a) 4.3% and (b) 3.0%.

The comparison between **Figure 9 (a)** and **(b)** essentially shows no difference by supplying additional Na. The corresponding detailed device statistics, shown in **Figure S15**, further reveal that there seems to be no significant gain by introducing NaF in our samples. The role of Na has been extensively discussed in both CIGS and CZTSSe devices. For CZTSSe, it is known that for Na-deficient substrates, as little as 5-10 nm of NaF can already produce significant improvements in the absorber layer morphology and device efficiencies, and above 20-30 nm NaF a deterioration of the device performance occurs, despite notable increases in grain sizes.[80–82] Therefore, our chosen value of 10 nm NaF allows us to conclude that our reference devices are not Na limited, and it should not be possible to solve the double layer issues by simply supplying more Na. In any case, we note that from experiments growing the kesterite directly on glass substrates, shown in **Figure S16** we have seen a significant enhanced in grain size, resulting from the oversupply of Na directly from the SLG. Therefore, it is clear that Na is playing a role in the double layer dynamics during annealing. Another example of this important role comes from our experiments using monocrystalline Si substrates, as shown in **Figure 9 (b)**. In our group, we have developed and characterized monolithic CZTS/Si tandem solar cells, and found that using monocrystalline Si substrates, where no native supply of Na exists, the kesterite always exhibits a double layer, even for thicknesses as low as 300 nm, as can be seen in previous work.[59,60] Remarkably, on Si substrates the double layer thickness proportion is again about 33%/67%, consistent with our previous claim that the upper/bottom layer thickness proprortions are set by the total availability of Cu in the precursor and the formation of the $Cu_{2-x}S$ phases, in a self-regulating process. We found that all the previous discussion regarding the influence of the annealing parameters is also valid for growth on Si substrates, but the thickness threshold for the formation of the double layer is much lower on Si substrates, which we can clearly attribute to the absence of the fluxing behavior provided by Na.

Given the decoupling of the effect of Na achieved with CZTS grown on monocrystalline Si substrates, this serves as an ideal platform to better resolve the composition gradients formed during annealing. Therefore, we have performed RBS and SIMS measurements comparing the precursor and the kesterite after annealing. The results are shown in **Figure 10**. A schematic of the constituting layers of the samples is shown as an inset. We have discussed the purpose of each layer in previous work.[59,60] Briefly, a TiN interlayer was also used in this work as it improves the CZTS adhesion and prevents the intermixing between any elements from CZTS and Si, allowing a better depth resolution in the measurements. In the RBS spectra of **Figure 10 (a)** and **(c)**, we can get depth-resolved and quantitative estimates of the film compositions by fitting the RBS yield curve with an appropriate software (RUMP was used in our case), near the energy regions (channels) where the Cu, Zn, Sn and S elements are detected. However, given the proximity in atomic mass between Cu and Zn, the yield due to these elements overlaps and only the sum of the concentration of both elements can be estimated.

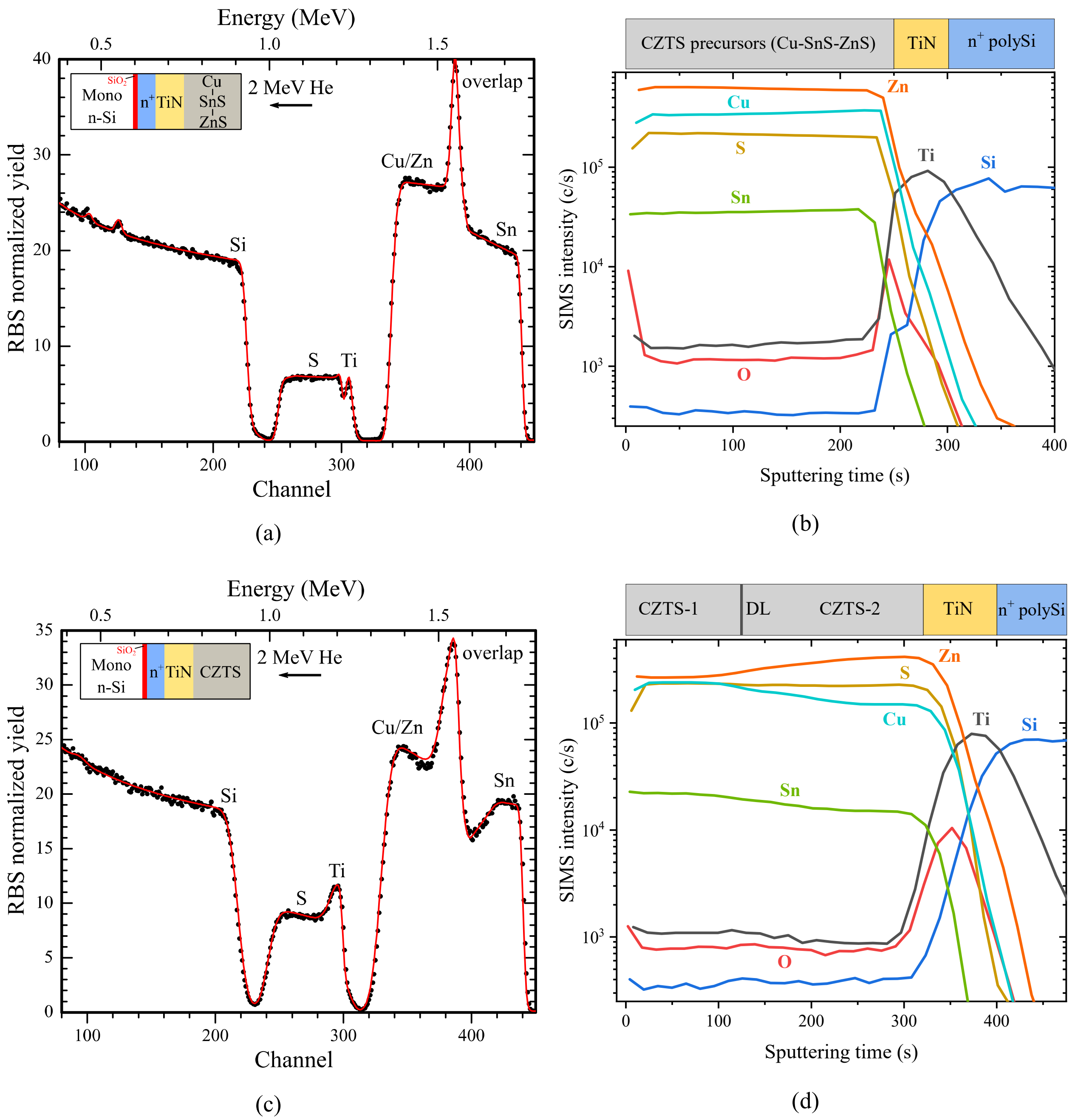

**Figure 10** – Depth profiling of CZTS grown on monocrystalline Si substrates with a 10 nm TiN adhesion layer. (a) and (b) are the RBS and SIMS profiles for the precursor, respectively, and (c) and (d) are the RBS and SIMS profiles for the kesterite after annealing the precursor in a 2S process. The label *overlap* in the RBS spectra decribes an overlap in the Cu, Zn and Sn signals. The red line in the RBS spectra is a fitting using the software RUMP.

For the cosputtered precursor in **Figure 10 (a)**, the yield from channels 330 to 450 indicates that the thickness-averaged precursor composition satisfies Cu/Sn + Zn/Sn = 3.63, which corresponds to slightly Cu-rich and Zn-rich conditions. Analyzing the sulfur content in channels 240 to 300 indicates that the S/metals ratio is approximately 0.65, which is consistent with our EDS measurements, and again shows that our cosputtered precursors are essentially sulfur deficient compared to the kesterite stoichiometry. As discussed in the previous sections, this sulfur deficiency could be fundamentally linked to the double layer occurrence. Furthermore, it also explains the incomplete reaction and heavy formation of secondary phases in the sample annealed without sulfur (**Figure S4**), and it is a well-known example of the Le Chatelier principle applied to kesterite synthesis.[83] Additionally, the increase in Sn signal from channels 450 to 390 allows us to estimate that the Sn concentration increases approximately linearly by about 13% from the surface of the precursor to the back interface. In contrast, the signal from channels 330 to 380 shows that the Cu+Zn concentration is approximately constant in depth. Cross-checking with the SIMS results for the precursor in **Figure 10 (b)** reveals that indeed there is an increase in the Sn signal towards the back. On the other hand, the Zn signal slightly decreases in depth, while the Cu signal increases, consistent with the approximately constant Cu + Zn concentration measured by RBS. These variations occur during the cosputtering process, but we do not have a clear understanding of the causes. These could possibly be related to a drift in the sputtering rate of the targets during the cosputtering process due to target heating or poisoning, or due to unintentional heating of the substrate itself during the process, resulting in an interdiffusion of the elements. However, the depth profiles are completely changed after annealing, as shown in **Figure 10 (c)** and **(d)**. In the RBS signal of **Figure 10 (c)**, in the region near channel 420, a sharp drop in the RBS yield for Sn can be seen. The fitting of this result is consistent with a model where the top 30% of the kesterite exhibits a nearly uniform composition of Cu/Sn + Zn/Sn = 3.3 and S/metals = 0.95, followed by a decrease in Sn concentration, which is estimated to be 40% lower in the back compared to the surface. The corresponding SIMS result of **Figure 10 (d)** matches this model, showing a nearly flat first third of the Cu, Zn and Sn, followed by a decrease in the Sn signal and again an opposite trend in the Cu and Zn signals. Remarkably, these are exactly the same trends as the SIMS result of **Figure 3 (c)**, which corresponds to a reference kesterite solar cell on Mo/SLG substrates with a median efficiency of 5.8%. Therefore, it appears as if these profiles are fundamentally a universal signature of the double layer dynamics in the kesterite. The approximate 30%/70% thickness proportions of the double layer (or of the composition gradients) also seem to appear sponteaneously in this case, without the supply of Na from the substrate. Therefore, this work proves that a fundamental redistribution of the metal elements occurs during annealing, which is brought about by the migration of the available Cu towards the surface, with the consequent formation of $Cu_{2-x}S$ phases at the surface, and the early establishement of highly asymmetric growth conditions that result in the composition gradients of the final kesterite absorbers. The exact numbers and metal ratios of the profiles given here will likely depend on the starting composition, although we note that our results are not far from the ideal Cu-poor/Zn-rich conditions.

As mentioned in the previous sections, a possible consequence of this asymmetric double layer growth is the occurrence of secondary phases outside those normally occurring due to the composition region of the kesterite. For an ideally Cu-poor and Zn-rich kesterite, this would mostly mean the segregation of a ZnS secondary phase. However, the possibly incomplete formation reaction associated with the asymmetry caused by the double layer formation might lead to the simultaneous presence of many types of secondary phases in the kesterite absorber. In **Figure S2 (b)**, using Raman spectroscopy we have determined that this can indeed be the case even in a device reaching an efficiency of 6.6%. In **Figure 11**, by using grazing incidence XRD, we again confirmed the existence of these secondary phases in the kesterite absorbers grown in this work on Mo/SLG and on Si substrates. A trace of monoclinic CTS is detected in the films, regardless of the substrate, in agreement with the Raman results of **Figure S2 (b)**. By tuning the incidence angle from 2° to 6°, thereby changing the depth of probing, we were able to further resolve the presence of the o-SnS phase (see **Figure S17**). Naturally, since all the kesterite films produced in this work are Zn-rich, the presence of the ZnS phase is very probable, although it cannot be easily discriminated using XRD. Still, this ZnS phase is visible in our Raman results (**Figure S2 (b)**).

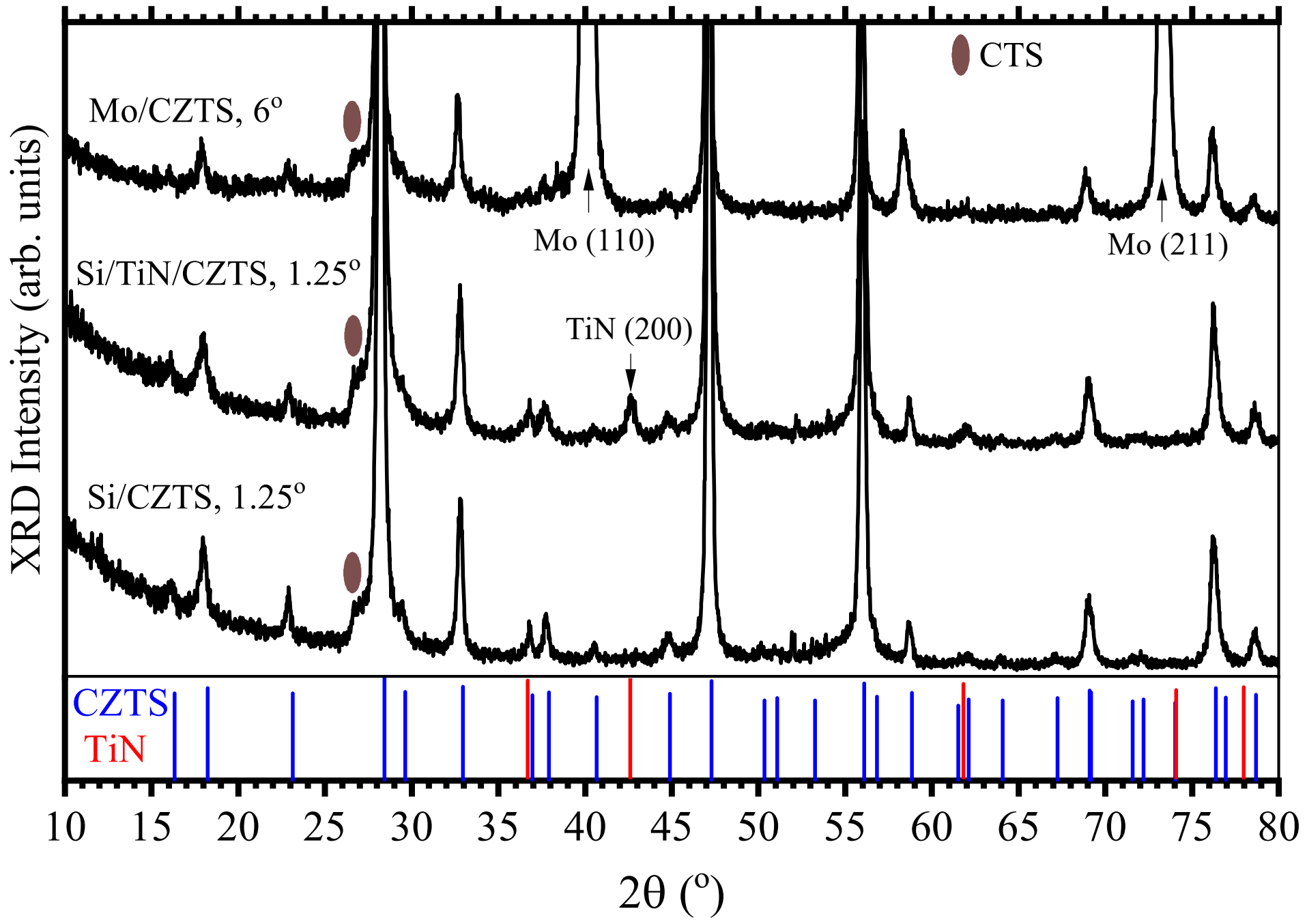

**Figure 11** – Grazing incidence XRD patterns for CZTS on Mo/SLG, CZTS on TiN/Si and CZTS on Si. The inset shows the reference positions of the peaks for CZTS (ICCD 04-015-0223) and for TiN (ICCD 01-087-0629), in a logarithmic scale for better visibility. For CTS, we used the JCPDS 04-010-5719 reference.

### *4.7 Comments on the generality of the double layer formation mechanism proposed in this work*

In this final section, we comment on the generality of our findings, and on possible comparisons with other kesterite synthesis methods. Regarding the first, in this discussion we have left out the possibility of tuning the cosputtering conditions. The only cosputtering parameters changed in this work were effectively the separate powers applied to each target (compositional tuning) and the cosputtering time (thickness control). This includes producing films with the same composition using significantly different power density regimes (the results of which are not discussed in this work for the sake of brevity). However, we acknowledge that tuning other parameters could have a great impact on the double layer dynamics. The notable mentions here are the sputtering pressure and the substrate temperature. The precursor films obtained in this work (examples of which can be seen in **Figure 2 (a)** and **(b)**) consist of amorphous, dense and highly reflective zone T films.[84] This is explained by a combination of a low substrate temperature (no intentional substrate heating was used in this work), relatively high melting point of the Cu, ZnS and SnS target materials (and thefore a low $T/T_m$, where $T$ is the substrate temperature and $T_m$ is the melting temperature of the coated material), and a low sputtering pressure ($4x10^{-3}$ mbar).[84] It is plausible that the persistent double layer formation could be associated with the resulting microstructure of our precursor films. For example, using substrate heating, Ren et al. reported columnar zone 2 precursor films from cosputtering of CuS, Zn and Sn targets, which resulted in single-layered CZTS absorbers.[78] In this work, we have tested the increase of the sputtering pressure from $4x10^{-3}$ to $1x10^{-2}$ mbar. The resulting kesterite morphology, which we show in **Figure S18**, still exhibited clear signs of the double layer segregation. However, substrate heating during cosputtering could have a larger influence on the double layer dynamics, and this approach is planned for future work.

Furthermore, even though the experimental part of this work focused exclusively on uniformly mixed, cosputtered precursors, a stacked configuration could provide some additional insights into the double layer dynamics. We have tested a Mo/ZnS/Cu/SnS configuration, which is shown in **Figure S19**. The resulting films exhibited a high density of voids near the middle, possibly related to a double layer behavior. Additionally, we observed some partial delamination issues using this route. Nevertheless, it is posssible that different stacking orders can lead to an improvement in the double layer occurrence. Given the results of our work, the position of the metallic Cu layer in the stack will presumably be of most importance in terms of the double layer behavior, although there are other factors to consider when optimizing a stacked system. Such a system has not yet been optimized in our group, but could be an interesting possibility for future work.

Considering our results and the detailed literature survey done in **Section 2**, some relevant observations can be made regarding the dependence of the double layer formation on the kesterite synthesis method. We note that, in the literature survey compiled in **Table 1**, all the sputtering-based approaches use a metallic Cu target, and in most cases a fully metallic precursor stack is used. That means that, in all cases, there is simultaneously the presence of metallic Cu and a sulfur deficiency in the respective precursor matrices. This appears to be consistent with our formation mechanism based on the superficial $Cu_{2-x}S$ appearance, although this is by no means the only parameter at play, as we thoroughly described in this work. Some successful examples of work not using Cu metal targets and with sulfur-rich precursors includes sputtering from a single quaternary target by Li et al.,[5] which resulted in a 11.95%-efficient CZTSe device, and cosputtering using $Cu_2S$ or CuS targets done by Ren and coworkers as mentioned in the previous paragraph.[15,30,78,85] In particular, besides using a Cu binary target, Ren and coworkers also used reactive sputtering with $H_2S$, ensuring a sulfur-rich precursor matrix. In these publications, there is no visible sign of a double layer formation at any point during the annealing stage. We can provide an additional example from other work in our group. We have also produced CZTS devices based on PLD of single targets, using oxide-based targets and sulfide-based targets.[42] For sulfide-based targets, the precursors consist of a mix of sulfide binaries generally with a high S/metals ratio (close to 1, in particular for low laser fluences), and indeed we do not see any evidence of double layer formation, even in absorbers with thicknesses above 2 μm, processed in the same 1S annealing conditions described in this work. An illustration of such an example can be found in **Figure S20**. Therefore, it is possible that the presence of $Cu_{2-x}$(S,Se) phases and high sulfur content in the precursor mix prior to the sulfurization/selenization annealing step plays a key role in determining the double-layer formation dynamics. This is also consistent with the improvements reported for the so-called "pre-annealing" steps in an inert atmosphere (no chalcogen available), as described in **Section 2**. There, the metallic Cu is partially converted into Cu(S,Se) within the precursor matrix. Nevertheless, this does not solve the sulfur deficiency limitation.

Another notable mention is cationic-substituted or alloyed kesterites. In particular, as we mentioned in **Section 2**, Giraldo et al. have convincingly shown that Ge alloying in the precursor matrix reduces the surface $Cu_{2-x}$(S,Se) dynamics and therefore significantly reduces the double layer.[21] Therefore, it is plausible that other cation substitutions can lead to a reduction in the double layer dynamics as well.

Nevertheless, considering the vast amount of publications where the double-layered kesterites appear, we believe that formation of this double layer should not be overlooked in the pursuit of paving the way for future device improvements.

## 5 Conclusion

In this work, we have reviewed the existing knowledge on double-layered kesterites, presented a detailed experimental study on the double-layer formation mechanisms and discussed their effect of the performance of sulfide kesterite devices. We show that for our cosputtering approach, the double layer is characterized by a large composition gradient, with Cu accumulation near the surface as a result of $Cu_{2-x}S$ phases forming early during the annealing. The resulting microstructure of the $Cu_{2-x}S$ layer likely acts as a sulfur barrier preventing a complete sulfurization of the precursor, in a self-regulating process, resulting in a double layer distribution with a constant top layer/bottom layer thickness proportion of about 33%/67%. Using CZTS samples grown on a monocrystalline Si substrate, we show that this thickness proportion and the respective composition gradients still occur, even without any Na or other alkali metal supply from the substrate. As a result, the kesterite bottom layer generally exhibits a morphology with smaller grain size and secondary phases, which we find to be independent of the overall kesterite composition, but instead a consequence of the incomplete conversion of the precursor into kesterite at the back side. By optimizing the annealing process in two separate annealing equipments, and tuning the relevant space of annealing parameters, we have validated our formation model and shown the universality of the double layer occurrence in our cosputtering system. Furthermore, we have shown that by decreasing the absorber thickness the double layer formation vanishes, yielding a significant improvement in the device parameters for lower absorber thicknesses, in particular the fill factor. Through a comparison with the well-known case of ultrathin CIGS devices, we concluded that the improvements are likely coming from the disappearance of the double layer in thin CZTS devices, which counterbalances the intrinsic losses occurring in thin absorbers.

Finally, we comment on the generality of the double layer formation mechanism proposed, by comparing our results to other synthesis routes, including results from solution processing and pulsed laser deposition processed in our group. We find that although the mechanism and exact experimental parameters for the double layer occurrence depend on the kesterite synthesis route, the common factors determining the double-layer occurrence appear to be the presence of metallic Cu and/or a chalcogen deficiency in the precursor matrix.

## Acknowledgments

This work was supported by a grant from the Innovation Fund Denmark (Grant 6154-00008A). F.M. thanks Stefanie Spiering at the Center for Solar Energy and Hydrogen Research (ZSW) for the help providing the Mo substrates.

## Associated Content

Supporting information available: further details on the double layer characterization can be consulted.

# TOC Graphic

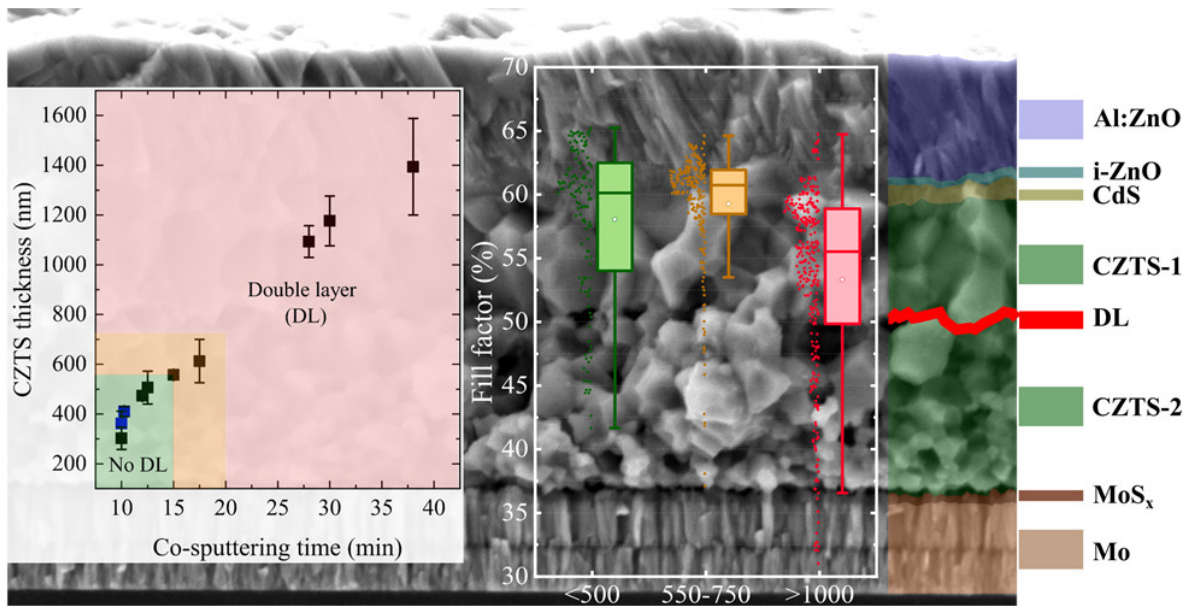